        \documentclass[%
        reprint,
        nofootinbib,
        amsmath,amssymb,
        aps,
        floatfix,
        ]{revtex4-2}
        \usepackage{cleveref}
        \usepackage{graphicx}
        \usepackage{dcolumn}
        \usepackage{bm}
        \usepackage{lipsum}
        
        \usepackage{wrapfig}
        \usepackage{mathrsfs}  
        
\begin{document}
        
        \preprint{APS/123-QED}
        
        \title{Single pion production in electron-proton interactions}
        
        \author{M. Kabirnezhad}
        \affiliation{%
          York University, Department of Physics and Astronomy, 4700 Keele Street, Toronto M3J 1P3, Canada\\
          University of Oxford, Denys Wilkinson Building, Oxford OX1 3RH, United Kingdom\\
          Imperial College London, Department of Physics,London SW7 2BZ, United Kingdom
        }
        \email{m.kabirnezhad@imperial.ac.uk}
        
        \date{\today}
        
        \begin{abstract}
          This paper presents an extension of the MK single pion production model \cite{MK, MK2} to high hadron invariant mass ($W$) and high momentum
          transfer ($Q^2$) to conform to the predictions of perturbative QCD due to the quark-hadron duality evidence. New form factors for several resonances and non-resonant background in the electron-nucleon cross-sections are determined taking into account the experimental data and improved evaluation techniques. 
          Fits to electron-proton scattering data are used to constrained free parameters and to assign the related uncertainties of the model. The results from this work can be used to determine the vector current in the corresponding neutrino-nucleon cross-sections, which is an important input for event generators in long baseline neutrino oscillation measurements.  
        \end{abstract}
        \maketitle
        
        \section{Introduction}
        Neutrino interactions that produce a single pion in the final state are of critical importance to accelerator-based neutrino experiments. Single pion production (SPP) channels make up the largest fraction of the inclusive neutrino-nucleus cross section in the $1-3 ~\text{GeV}$ region covered by most accelerator-based neutrino beams. Models of SPP cross section processes are required to accurately predict the properties of the final state (number of particles and their kinematics) in neutrino interactions. Full predictions incorporate models of the neutrino-nucleon interactions \cite{Adler,RS,Rein,sato,HNV,Mariano2011,DCC,Alam,Raul_nucleon,Luis_wpp,Luis_NC,proc1,proc2,proc3,MK,MK2,Sato21}, models of the nucleus \cite{RMF,Raul,Luis,tania,alexis,Sajjad} (in which the target nucleon resides, and which must be traversed by the interaction's final-state particles) and models of detector response.\\
        
        An accurate model for the neutrino-nucleon reaction is essential for an accurate description of neutrino-nucleus interactions. However, a nuclear model is still required to perform comparisons with neutrino-nucleus data. Model comparisons can only test the convolution of nucleon level plus nuclear models. However, if the nucleon level model is known to be accurate, comparison can be used to infer information about the suitability of the nuclear model.\\
        
        The underlying fundamental theory of SPP from a nucleon is quantum chromodynamics (QCD), and QCD calculations provide important and reliable predictions in the perturbative (high-$Q^2$) domain. However, in the resonance region (low-$Q^2$, non-perturbative domain), phenomenological models are necessary to describe SPP. Fortunately, quark-hadron duality \cite{FF_QCD,lalak2008,Jorge_SIS} (which establishes a relationship between the quark-gluon description at high $Q^2$, and the hadron-meson description at low $Q^2$) allows for the use of perturbative QCD calculations to constrain the asymptotic (high-$Q^2$) behaviour of form factors in the phenomenological models. Incorporating duality into the model building ensures that the phenomenological constraints from data tend towards the QCD predictions as $Q^2$ increases. \\
        
        Most phenomenological models are developed in the low $Q^2$ region ($Q^2 < 1.5~  (\text{GeV}/c)^2$) without using the perturbative QCD constraints of the high $Q^2$ limit. They include a description of single pion production through two distinct channels: \emph{i) resonance production} is where the exchange boson has the requisite four-momentum to excite the target nucleon to a resonance state, which then promptly decays to produce a final-state meson, and \emph{ii) non-resonant production}, where the pion is created at the interaction vertex (with the outgoing nucleon), which produces final states identical to resonance production that results in non-negligible interference terms. \\
        
        The majority of SPP models are in quantitative agreement with the experimental data in the first resonance region ($W<1.3~ \text{GeV}$)  where the resonance structure and the decay amplitude of the $\Delta$ resonance are well characterised. However, understanding SPP in the second resonance and the third resonance regions (higher $W$), where several resonances overlap and interfere with each other, is crucial for neutrino oscillation experiments such as No$\nu$A and DUNE. The most onerous challenge in developing a high precision SPP neutrino interaction model is to correctly include the effects of overlapping structures from the many resonances, each with their own form-factors, that contribute to the hadronic tensor. \\
        
        SPP models must contain adjustable parameters associated with each resonance form factors, which can be constrained by data. These parameters from the vector current can be fit to existing electron scattering data, as discussed in Ref.~\cite {MK2}. However, in this paper, the  model is extended to a region with higher $Q^2$ and $W$, known as the transition region between resonance and deep inelastic scattering (DIS) regions (sec.~ \ref{update}). Therefore all available electron scattering SPP data are included in the analysis. Also, the analysis and the evaluation of the systematic uncertainties is improved, as discussed in sec.~\ref{analys}.\\ 
        \section{The update of the MK model \label{update}}
        
        The first iteration of the MK model \cite{MK,MK2} provides a full kinematic description of SPP in neutrino-nucleon interactions, including the resonant and the non-resonant interactions plus their interference terms in the helicity basis. In this framework, the cross-section is defined (Eq. 18 of Ref. \cite{MK}) from the summation of all helicity amplitudes of resonances and non-resonant backgrounds. \\
        
        The helicity amplitudes are defined by the incident nucleon helicity ($\lambda_1$), the outgoing nucleon helicity ($\lambda_2$), and the gauge boson's polarisation ($\lambda_k$):
        \begin{eqnarray}\label{HA_definition}
        \tilde{F}_{\lambda_2, \lambda_1}^{\lambda_k} = \langle~N\pi|~e^{\rho}_{\lambda_k} (\frac{1}{2M}) J^V_{\rho}~|N ~\rangle
        \end{eqnarray}
        where $J^V$ is the hadronic vector current and $e_{\lambda_k}$ ($e_L$, $e_R$, $e_{\pm}$) are polarisation of the gauge boson (transverse and longitudinal) vectors:
        \begin{eqnarray}\label{pol_vec}
        e^{\alpha}_{L} &=& \frac{1}{\sqrt{2}} \begin{pmatrix} 0&1 &-i&0\end{pmatrix} ~,\nonumber\\
        e^{\alpha}_{R} &=& \frac{1}{\sqrt{2}} \begin{pmatrix} 0&-1&-i&0 \end{pmatrix}~, \nonumber\\
        e^{\alpha}_{\lambda} &=& \frac{1}{\sqrt{|(\epsilon^0_{\lambda})^2 - (\epsilon^3_{\lambda})^2 |}}  \begin{pmatrix} \epsilon^0_{\lambda}&0&0&\epsilon^3_{\lambda}\end{pmatrix} \label{W_pol}\nonumber
        \end{eqnarray}
        where $\lambda = \mp$ and  $\epsilon^{0}_{\lambda}$ ($\epsilon^{3}_{\lambda}$) is the first (last) component of the lepton currents:
        \begin{equation}
        \epsilon_{\lambda}^{\rho}= \bar{u}_{l' _{\lambda}}(k_2) \gamma^{\rho}(1- \gamma_5) u_{l }(k_1). \label{epsilonLR}
        \end{equation}
        The explicit expressions of the helicity amplitude for resonances and non-resonant backgrounds \cite{MK,MK2} are reproduced in Appendixes \ref{appA} and \ref{appB}. The helicity amplitudes of the resonances are a product of the resonance production amplitudes ($f^V_{-1}, f^V_{-3}$ and $f^{V(\lambda)}_{0+}$) and the decay amplitudes. Resonance production in the MK model \cite{MK2} follows the Rarita-Schwinger formalism \cite{Olga_fit} for the first and the second resonance regions ($W<1.6~ \text{GeV}$) and the formalisms of the Rein-Sehgal model \cite{RS} in the third region ($W>1.6~ \text{GeV}$). The non-resonant interactions follow the Hernandez {\it et al.} model \cite{HNV} which is deduced from the chiral-perturbation (ChPT) theory Lagrangian of the pion-nucleon system and it is not reliable at high hadron invariant mass ($W>1.4$ GeV). The vector form-factors were defined to improve agreement with exclusive electron-scattering data in the resonance region ($Q^2 < 1 ~\text{GeV}/c)^2$.\\
        
        In the updated approach presented here, the MK model is extended to higher momentum transferred ($Q^2$) by introducing form factors based on vector meson dominance (VMD) models consistent with the Quantum Chromodynamics (QCD) theory. The form factors should exhibit a certain asymptotic $Q^2$ behaviour to satisfy the asymptotic condition for helicity amplitudes at $Q^2 \rightarrow \infty$, as prescribed by perturbative QCD (pQCD) \cite{FF_QCD,lalak2008}. \\
        
        The VMD model is based on the strongly interacting virtual vector mesons as intermediaries in the coupling between a virtual photon and a nucleon. The model has been a successful theory to determine form factors\footnote{The dipole form factor would be obtained if vector mesons propagate between the virtual photon and the nucleon.} at low $Q^2$ and can reproduce $Q^2$-evolution of form-factors at high $Q^2$, to join smoothly with pQCD calculations \cite{stoler}. Thus, the VMD models being consistent with pQCD-predictions show a satisfactory result in both perturbative and non-perturbative domains \cite{GK,FF_QCD}. A list of vector mesons \cite{FF_QCD} is shown in \Cref{meson}. \\
        
        \begin{table}
          \centering
          \caption{vector- meson masses}
          \label{meson}
          \renewcommand{\arraystretch}{1.3}
          \begin{ruledtabular}
            \begin{tabular}{lcccl}
              $k$&$ \rho$ -group  & $m_{(\rho)k}[\text{GeV}]$ &$\omega$ -group&  $m_{(\omega)k}[\text{GeV}]$ 
              \\ [0.1ex]
              \hline
              1&$\rho(770)$ & 0.77526 & $\omega(782)$& 0.78265\\
              2&$\rho(1450)$ & 1.465 & $\omega(1420)$& 1.410\\
              3&$\rho(1700)$ & 1.720 & $\omega(1650)$& 1.670\\
              4&$\rho(1900)$ & 1.885 & $\omega(1960)$& 1.960\\
              5&$\rho(2150)$ & 2.149 & $\omega(2150)$& 2.148\\
            \end{tabular}
          \end{ruledtabular}
        \end{table}
        
        \subsection{Resonant interactions}
        
         The resonance production amplitudes, $f^V_{-1}, f^V_{-3}$ and $f^{V(\lambda)}_{0+}$, contain the transition nucleon form-factors. They are presented in the previous paper \cite{MK2} and are used in this work to incorporate as the electron scattering data is used in this work too. When used in neutrino interactions, the mass of the outgoing charged lepton can not be ignored for charge current interactions. Therefore, the more general vector helicity amplitudes for the $P_{33}(1232)$, $P_{11}(1440)$, $D_{13}(1520)$, and $S_{11}(1535)$ resonances, which can be used for both electron and neutrino interactions, are presented\footnote{For the rest of the resonances, the helicity amplitudes in the Rein-Sehgal model with the lepton mass correction are used \cite{MK, MK2}}  in Eqs.~(\ref{delta}) - (\ref{last}) (Note that for consistency all the notations in the following relations are taken from the previous works \cite{MK,MK2}, such as the nucleon's mass ($M$), the invariant mass ($W$), $W_{\pm}= W \pm M$, $k=k_1- k_2$, where $k_{1(2)}$ is the momentum of the incoming (outgoing) lepton, $Q^2=-k^2 = -(k_1 -k_2)^2= {\bf{k}}^2 - k_0^2$, and $E_k=\sqrt{M^2 + {\bf{k}} ^2}$.):
        \begin{itemize}
        \item{resonance $P_{33}(1232)$}
          \begin{eqnarray}
            f^{V}_{-3} &= &-\frac{|{\bf{k}}|}{\sqrt{2W(E_k +M)} }\bigg[\frac{C_3 W_{+}}{M} + \frac{C_4}{M^2} Wk_0 \nonumber\\
                       &+&  \frac{C_5}{M^2} (Wk_0 -k^2 )\bigg],\nonumber\\
            f^{V}_{-1} &= &\frac{|{\bf{k}}|}{\sqrt{6W(E_k +M)} }\bigg[\frac{C_3}{MW}\left(k^2 - MW_{+}\right) \nonumber\\
                       &+& \frac{C_4}{M^2} Wk_0 +  \frac{C_5}{M^2} (Wk_0 -k^2)\bigg],\nonumber\\
            f^{V(\lambda)}_{0+} &= &-\frac{1}{C_{\lambda}}\frac{|{\bf{k}}|}{\sqrt{3W(E_k +M)} } \left(|{\bf{k}}|\epsilon^0_{\lambda} - k_0\epsilon^3_{\lambda} \right) \bigg[\frac{C_3}{M} \nonumber\\
                       &+& \frac{C_4}{M^2}W  +  \frac{C_5}{M^2} (W- k_0)\bigg]\nonumber\\
                       \label{delta}
          \end{eqnarray} 
        \item{resonance $D_{13}(1520)$}
          \begin{eqnarray}
            f^{V}_{-3} &= &\sqrt{\frac{E_k + M}{2W}}\bigg[\frac{C_3}{M}W_{-} + \frac{C_4}{M^2} Wk_0
                            \frac{C_5}{M^2} \left(Wk_0 - k^2\right)\bigg],\nonumber\\
            f^{V}_{-1} &= &\sqrt{\frac{E_k + M}{6W}}\bigg[\frac{C_3}{M}\left(W _{-}- 2\frac{{\bf{k}}^2}{E_k + M} \right) \nonumber\\
                       &+ &\frac{C_4}{M^2}Wk_0  +  \frac{C_5}{M^2}\left(Wk_0 - k^2\right)  \bigg],\nonumber\\
            f^{V(\lambda)}_{0+} &= &\frac{1}{C_{\lambda}} \sqrt{\frac{E_k + M}{3W}}\left(|{\bf{k}}|\epsilon^0_{\lambda} - k_0\epsilon^3_{\lambda} \right) \bigg[\frac{C_3}{M}  + \frac{C_4}{M^2}W\nonumber\\
                       & +&   \frac{C_5}{M^2} (W- k_0)\bigg], \nonumber\\
          \end{eqnarray} 
        \item{resonance $P_{11}(1440)$ }
          \begin{eqnarray}
            f^{V}_{-1} &= &\frac{|{\bf{k}}|}{\sqrt{W(E_k +M)}} \left[g_1 - \frac{g_2}{W_{+}^2} k^2\right], \nonumber\\
            f^{V(\lambda)}_{0+} &= &-\frac{1}{C_{\lambda}}\frac{|{\bf{k}}|}{\sqrt{2W(E_k +M)} } \left(|{\bf{k}}|\epsilon^0_{\lambda} - k_0\epsilon^3_{\lambda} \right) \frac{1}{W_{+}}\left[ g_1 - g_2\right],\nonumber\\
          \end{eqnarray} 
        \item{resonance $S_{11}(1535)$}
          \begin{eqnarray}
            f^{V}_{-1} &= &\sqrt{\frac{(E_k + M)}{W}}\left[\frac{g_1}{W_{+}^2} k^2 - \frac{g_2}{W_{+}}W_{-} \right], \nonumber\\
            f^{V(\lambda)}_{0+} &= &\frac{1}{C^V_{\lambda}} \sqrt{\frac{E_k + M}{2W}}\left(|{\bf{k}}|\epsilon^0_{\lambda} - k_0\epsilon^3_{\lambda} \right) \left[-\frac{g_1W_{-}}{W_{+}^2} + \frac{g_2}{W_{+}}\right], \nonumber\\
        \label{last}  \end{eqnarray} 
        \end{itemize}
        where the lepton mass appears in $C_{\lambda}=\sqrt{|(\epsilon^0_{\lambda})^2 - (\epsilon^3_{\lambda})^2 |}$.  $C^V_3$, $C^V_4$, and $C^V_5$ ($g^V_1$ and $g^V_2$) are the VMD form-factors for resonances with spin-3/2 (spin-1/2). Using the parametrisation suggested by Vereshkov and  Volchanskiy \cite{FF_QCD} yields:
        \begin{eqnarray} \label{FFCg}
          C_{\alpha}^{(p)}(k^2)&=& \frac{C_{\alpha} ^{(p)}(0)}{L_{\alpha}^{(p)}(k^2)}\sum_{k=1}^{K} \frac{a_{\alpha k }m^2_k}{m^2_k- k^2} ~~~(\alpha=3-5), \label{3half} \\
          g_{\beta}^{(p)}(k^2) &=&\frac{g_{\beta}^{(p)}(0)}{L_{\beta}^{(p)}(k^2)}\sum_{k=1}^{K} \frac{b_{\beta k }m^2_k}{m^2_k - k^2} ~~~(\beta=1-2). \label{1half}
        \end{eqnarray} 
        Here $m^2_k= m^2_{(\rho)k}$ for the $\Delta$ ($P_{33}(1232)$) resonance and $m^2_k= (m^2_{(\omega)k} + m^2_{(\rho)k})/2$ for the resonances in the second region: $P_{11}(1440)$, $D_{13}(1520)$, and $S_{11}(1535)$. \Cref{meson} gives the masses $m_{(\rho)k}$ and $m_{(\omega)k}$. 
        The logarithmic renormalisation $L_{\alpha}^{(p)}(k^2)$ and $L_{\beta}^{(p)}(k^2)$ are defined to retain the asymptotic behaviour as it is predicted by pQCD \cite{FF_QCD}:
        \begin{eqnarray} \label{log}
          L_{\alpha (\beta)}^{(p)}(k^2) = &\Bigg[&1 + h_{\alpha (\beta)}^{(V)}\ln \left(1 - \frac{k^2}{\Lambda^2_{QCD}}\right)\nonumber\\ &+& k_{\alpha (\beta)}^{(V)}\ln^2 \left(1 - \frac{k^2}{\Lambda^2_{QCD}}\right)  				\Bigg]^{n_{\alpha(\beta)}/2},
        \end{eqnarray} 
        where $n_1 = n_3 \simeq 3$, $n_1>n_2$ and $n_5>n_3>n_4$.\\
        
        The parameters $a_{\alpha k }$ and $b_{\beta k }$ must obey a number of relations to fulfill asymptotic QCD (quark-hadron duality). They are referred to as \emph{linear superconvergence} relations as discussed in Ref.~\cite{FF_QCD}.\\ 
        
        For $\alpha=3$,
        \begin{eqnarray} \label{converg1}
          &\sum_{k=1}^{K}& a_{3 k }=1,~~~~~~~~~~~~~~~\sum_{k=1}^{K} a_{3 k } m^2_{k}=0,\nonumber\\ 
          &\sum_{k=1}^{K}& a_{3 k } m^4_{k}=0.
        \end{eqnarray} 
        
        For $\alpha=4$ and $5$,
        \begin{eqnarray} \label{converg2}
          &\sum_{k=1}^{K}& a_{\alpha k }=1,~~~~~~~~~~~~~~\sum_{k=1}^{K} a_{\alpha k } m^2_{k}=0,\nonumber\\ 
          &\sum_{k=1}^{K}& a_{\alpha k } m^4_{k}=0,~~~~~~~~~~\sum_{k=1}^{K} a_{\alpha k }m^6_{k}=0.
        \end{eqnarray} 
        
        For $\beta=1$,
        \begin{eqnarray} \label{converg3}
          &\sum_{k=1}^{K}& b_{1 k }=1,~~~~~~~~~~~~~~~\sum_{k=1}^{K} b_{3 k } m^2_{k}=0,\nonumber\\ 
          &\sum_{k=1}^{K}& b_{1 k }m^4_{k}=0.
        \end{eqnarray} 
        For $\beta=2$,
        \begin{eqnarray} \label{converg4}
          &\sum_{k=1}^{K}& b_{2 k }=1,~~~~~~~~~~~~~~\sum_{k=1}^{K} b_{2 k } m^2_{k}=0,\nonumber\\ 
          &\sum_{k=1}^{K}& b_{2 k }m^4_{k}=0,~~~~~~~~~~\sum_{k=1}^{K} b_{2 k } m^6_{k}=0.
        \end{eqnarray} 
        \\
        \subsection{Non-resonant background} \label{nonres_VMD}
        The helicity amplitudes of non-resonant background from the chiral-perturbation (ChPT) diagrams were calculated in the previous work \cite{MK} and are reproduced in Appendix \ref{appB}. In this work, the dipole vector form factors of ChPT theory (low W) are changed to VMD form-factors as suggested by Lomon \cite{GK}. They are 
        \begin{widetext}
          \begin{eqnarray} \label{GK1}
            F_1^{iv} (k^2)&= &\frac{N}{2} \frac{1.0317 + 0.0875(1 - k^2/0.3176)^{-2}}{1 - k^2/0.5496} F_1^{\rho}(k^2) +  \mathcal{R}_{\rho'} \frac{m_{\rho'}^2}{m_{\rho'}^2 - k^2 }F_1^{\rho}(k^2) + (1 - 1.1192N/2 - \mathcal{R}_{\rho'}) 	F_1^{D}(k^2),\nonumber\\
            F_2^{iv} (k^2)&= &\frac{N}{2} \frac{5.7824 + 0.3907(1 - k^2/0.1422)^{-1}}{1 - k^2/0.5362} F_2^{\rho}(k^2) + \kappa_{\rho'}\mathcal{R}_{\rho'}\frac{m_{\rho'}^2}{m_{\rho'}^2 - k^2 }F_2^{\rho}(k^2) + (\kappa_v - 6.1731N/2 - \kappa_{\rho'}\mathcal{R}_{\rho'}) F_2^{D}(k^2),\nonumber\\
            F_1^{is} (k^2)&= &\mathcal{R}_{\omega}\frac{m_{\omega}^2}{m_{\omega}^2 - k^2 }F_1^{\omega}(k^2) +    \mathcal{R}_{\phi}\frac{m_{\phi}^2}{m_{\phi}^2 - k^2 }F_1^{\phi}(k^2)  + (1 - \mathcal{R}_{\omega} ) F_1^{D}(k^2),\nonumber\\
            F_2^{is} (k^2)&= &\kappa_{\omega}\mathcal{R}_{\omega}\frac{m_{\omega}^2}{m_{\omega}^2 - k^2 }F_2^{\omega}(k^2)  +  \kappa_{\phi}\mathcal{R}_{\phi}\frac{m_{\phi}^2}{m_{\phi}^2 - k^2 }F_2^{\phi}(k^2) (\kappa_s - \kappa_{\omega}\mathcal{R}_{\omega}  - \kappa_{\phi}\mathcal{R}_{\phi}) F_2^{D}(k^2),
          \end{eqnarray} 
        \end{widetext}
        where $\kappa_v=3.793$, $\kappa_s=-0.12$, $m_{\rho'}=1.465~\text{GeV}$, $m_{\omega}=0.78265~\text{GeV}$ and $m_{\phi}=1.020~\text{GeV}$. The vector form-factors, $ F_{1,2}^{\rho}$, $ F_{1,2}^{\omega}$ and $ F_{1,2}^{D}$, are given by:
        \begin{eqnarray}\label{GKFF1}
          F_1^{\mathcal{M},D}(k^2)&=& \frac{\Lambda^2_{1,D}}{\Lambda^2_{1,D} - \tilde{k}^2} \frac{\Lambda^2_{2}}{\Lambda^2_{2} - \tilde{k}^2}, \nonumber\\
          F_2^{\mathcal{M},D}(k^2)&=& \frac{\Lambda^2_{1,D}}{\Lambda^2_{1,D} - \tilde{k}^2} \left(\frac{\Lambda^2_{2}}{\Lambda^2_{2} - \tilde{k}^2}\right)^2,
        \end{eqnarray}
        where $\mathcal{M}= \rho$ and $\omega$ and $\Lambda_{1,D}$ is $\Lambda_1$ for $F_{1,2}^{\mathcal{M}}$ and $\Lambda_D$ for $F_{1,2}^{D}$,
        \begin{eqnarray} \label{GKFF2}
          F_1^{\phi}(k^2)&=& F_1^{\mathcal{M}} \left(\frac{-k^2}{\Lambda^2_{1} - k^2}\right)^{3/2} ,~~~~ F^{\phi}_1(0)=0, \nonumber\\
          F_2^{\phi}(k^2)&=& F_2^{\mathcal{M}} \left(\frac{\mu^2_{\phi} - \Lambda_1^2 k^2}{\mu^2_{\phi}\Lambda^2_{1} - k^2}\right)^{3/2},
        \end{eqnarray}
        where
        \begin{eqnarray}
          \tilde{k}^2 = k^2 \frac{\ln\left[(\Lambda^2_D - k^2)/\Lambda_{QCD}^2 \right]}{\ln(\Lambda^2_D/\Lambda^2_{QCD})}~~. \nonumber 
        \end{eqnarray}
        
        The non-resonant background of the MK model is extended to high $W$ ($W>1.4~\text{GeV}$), using the Regge trajectory approach \cite{guidal}. The propagators of the \emph{t}-channel meson exchange ChPT diagrams are replaced by the corresponding Regge propagators (ReChi). The expressions of the Regge propagators for the meson trajectories of the $\pi$ and $\rho$ are the following:
        \begin{eqnarray}\label{regge}
          \mathcal{P}_{\pi}(t,s) &=& -\alpha'_{\pi} \Gamma(- \alpha_{\pi}(t) )\left(\frac{S}{S^{\pi}_0}\right)^{\alpha_{\pi}(t)}, \nonumber\\
          \mathcal{P}_{\rho}(t,s) &=& -\alpha'_{\rho} \Gamma(- \alpha_{\rho}(t) )\left(\frac{S}{S^{\rho}_0}\right)^{(\alpha_{\rho}(t) - 1)},
        \end{eqnarray}
        where $S^{\pi}_0$ and $S^{\rho}_0$ are adjustable parameters. The Regge trajectory of $\pi$ and $\rho$ are as follows:
        \begin{eqnarray}\label{trajectory}
          \alpha_{\pi}(t) &=& 0.75(t- m_\pi^2),\nonumber\\
          \alpha_{\rho}(t) &=& 0.53 + 0.85t.
        \end{eqnarray}
        Following the Hybrid model's parametrisation \cite{hybrid}, the helicity amplitudes for non-resonant background will be:
        \begin{eqnarray}\label{hyb}
        \tilde{F} = \cos^2 \phi(W) \tilde{F}_{ChPT} + \sin^2 \phi(W)\tilde{F}_{ReChi},
        \end{eqnarray}
        where  $\tilde{F}= \tilde{F}_{\lambda_2, \lambda_1}^{\lambda_k}$ and 
        \begin{eqnarray}\label{hyb2}
         \phi(W) = \frac{\pi}{2} \left( 1- \frac{1}{\frac{\exp (W-1.7)}{0.1}} \right).
        \end{eqnarray}
        $\tilde{F}_{ChPT}$ are the helicity amplitudes for the ChPT background and $\tilde{F}_{ReChi}$ are the modified helicity amplitudes with Regge propagators.\\
        
        This approach with the linear Regge trajectories (Eq.~\ref{trajectory}) is valid at high-energy and low-momentum transfer. To ensure that the model reproduces data at high momentum transfer, the trajectories are multiplied by a factor of $(1 + 2.4Q^2/W^2)^{-1}$, as suggested in Ref.~\cite{cutoff}. The adjustable parameters in the Regge trajectory and VMD form factors are fit to the available exclusive electron scattering data in the next section.\\ 
        
        \section{Analysis of electron-induced exclusive data\label{analys}}
        Exclusive SPP electron-proton scattering data are used to fit the relevant free parameters of the model. Fits were used to determine the $Q^2$ dependence of the transition form-factors for resonance production and non-resonant background SPP. 
        The analysis method is similar to the previous work \cite{MK2}. There measurements of the single pion differential cross sections for electrons scattering off of a hydrogen target in a limited kinematic region \cite{pi0,pi+} were used in the fits. However, in this analysis, all available data from the CLAS Collaboration \cite{pi0,pi+,clas_lowQ,pi0_high,pi+_high}  in the relevant kinematic region ($Q^{2}\in[0.16-6.00]~(\text{GeV}/c)^2$ and $W\in[1.1-2.01]~\text{GeV}$) were used.\\
        
        A list of resonances with vector currents, used in this updated MK model, is given in \Cref{res_list}. All resonances are included since data with higher $W$ are included in the fit. Therefore, all resonances in \Cref{res_list} will have their own form-factors. This differs from the previous analysis where only data with $W<1.68~\text{GeV}$ were included in the fit.\\
        \begin{table}
          \centering
          \caption{Nucleon-resonances, resonance's mass ($M_R$), resonance's width ($\Gamma_0$), branching ratio ($\chi_E$), resonance's signs ($\sigma^D$), and resonance's phases.   }
          \label{res_list}
          \renewcommand{\arraystretch}{1.3}
          \begin{ruledtabular}
            \begin{tabular}{lccccl}
              Resonance & $M_R(\text{MeV})$ &$\Gamma_0(\text{MeV})$& $\chi_E$&$ \sigma^D$ &Phase  
              \\ [0.1ex]
              \hline
              $P_{33}(1232)$ & 1236 & 130& 0.994   & + & 0.55\\
              $ P_{11}(1440)$ & 1410 & 423& 0.65& +  &0.0\\
              $D_{13}(1520)$ &1519  & 120& 0.60& + &1.68\\
              $S_{11}(1535)$ & 1545&175 &  0.45& -  &1.60\\
              $P_{33}(1600)$ & 1.64 & 200& 0.16   & +&1.98\\
              $S_{31}(1620)$ & 1610 &130& 0.3 & +&--- \\
              $F_{15}(1680)$&1685&120 & 0.65  & - &--- \\
              $D_{33}(1700)$&1710&300& 0.15   & + &--- \\     
              $  P_{11}(1710)$&1710&140& 0.11   & -  &---\\
              $  P_{13}(1720)$ & 1720&250& 0.11 & + &--- \\      
              $ F_{35} (1905)$ & 1880&330& 0.12 & -  &---\\
              $ P_{31} (1910)$ & 1900&300 &0.22 & + &---\\
              $ P_{33}(1920)$ &  1920&300 & 0.12& - &---\\
              $ F_{37}(1950)$ & 1930&285 &0.40  & + &---\\
            \end{tabular}
          \end{ruledtabular}
        \end{table}
        
        The use of external data to constrain the free parameters of the model has a large impact on the model predictions. It is therefore crucial to minimise the bias that would come from a single experiment's measurement via systematic uncertainties or by extrapolating data outside the regions of that experiment's kinematic coverage. A simultaneous fit to the full dataset properly accounts for correlations in the experimental data. The evaluation provides a covariance matrix for the fit parameters required to fully characterise the systematic uncertainty of the model. By design this matrix properly includes correlations in the data as well as those enforced by the model.\\
           
        As a practical matter, the addition of several datasets from different channels and kinematic regions can lead to Peelle's Pertinent puzzle \cite{ppp}. It is likely to occur when a significant discrepancy exists or some experimental uncertainties are unaccounted for or underestimated. As described in Ref.~\cite{jlab_database}, some of the electron scattering data used in this fit do not have systematic uncertainties evaluated for all angular bins, and in some cases, no systematic uncertainties are reported.
        In addition, the measurements (see \Cref{clas_data}) for the $e p \rightarrow e n \pi^{+}$  channel at high $Q^2$ (or $W$) have uncertainties lower than those for measurements of other channels. This result tends to favour the data used to constrain these channels noticeably over the $e p \rightarrow e p \pi^{0}$ channel. 
        Therefore, weighting factors are introduced for problematic datasets in order to avoid Peelle's Pertinent puzzle and to reduce any bias to regions not covered by those data. The weighting factors for different datasets used in this analysis are  shown in \Cref{clas_data}. \\
        \begin{table}
          \centering
          \caption{Datasets analysed by the CLAS Collaboration over a vast kinematic range for two channels and their weighting factors (WF) used in this analysis.}
          \label{clas_data}
          \renewcommand{\arraystretch}{1.3}
          \begin{ruledtabular}
            \begin{tabular}{lcccl}
              Channel & $E_e~\text{(GeV)}$&$Q^2~(\text{GeV}/c)^2$& $W~\text{(GeV)}$&WF 
              \\ [0.1ex]
              \hline
              $e p \rightarrow e p \pi^{0}$ & 1.046  & 0.16 - 0.32  & 1.1 - 1.34 & 0.85\\
              $e p \rightarrow e n \pi^{+}$ & 1.046  & 0.16 - 0.32  & 1.1 - 1.34& 0.5\\
              $e p \rightarrow e n \pi^{+}$ & 1.515  & 0.30 - 0.60 & 1.11 - 1.57& 1\\
              $e p \rightarrow e p \pi^{0}$ & 1.645  & 0.40 - 0.90  & 1.1 - 1.68& 1\\
              $e p \rightarrow e p \pi^{0}$ & 2.445  & 0.65 - 1.80  & 1.1 - 1.68& 1\\
              $e p \rightarrow e n \pi^{+}$ & 5.499  & 1.80 - 4.00 & 1.62 - 2.01& 0.5\\
              $e p \rightarrow e n \pi^{+}$ & 5.754  & 1.72 - 4.16 & 1.15 - 1.67& 0.5\\
              $e p \rightarrow e p \pi^{0}$ & 5.754  & 3.00 - 6.00 & 1.11 - 1.39& 1\\
            \end{tabular}
          \end{ruledtabular}
        \end{table}
        
        The $88$ free parameters cover all $14$ transition form factors for resonance production and non-resonant backgrounds. These parameters are fit to the experimental data using the MIGRAD algorithm in the MINUIT2 software package. The best-fit results with a reduced $\chi^2$ of $3.05$ are discussed below.
        \subsection{VMD form-factors for spin $3/2$ resonances}
        There are two spin-$3/2$ resonances in the first and the second resonance regions, i.e. $\Delta$ ($P_{33}(1232)$) and $D_{13}(1520)$, with the form-factors given in Eq.~(\ref{3half}). According to Ref.~\cite{FF_QCD}, the first four mesons in \Cref{meson} are enough to describe the VMD form factors of the spin-$3/2$ resonances. Therefore, $K=4$ in Eq.~(\ref{3half}). All the parameters in $a_{\alpha k}$, $\alpha = 4$ and $5$, are fixed by the four superconvergence relations in Eq.~(\ref{converg2}). However the parameters $a_{3 k}$ satisfy three superconvergence relations [Eq.~(\ref{converg1})] and only one free parameter remains. The QCD-scale in the logarithmic renormalisations [Eq. (\ref{log})] can vary between $\Lambda_{QCD} \in [0.19-0.24]~\text{GeV}$ \cite{FF_QCD} and  $n_{\alpha=3}=3$, $n_{\alpha=4}=2$ and $n_{\alpha=5}=4$. Also, a cut-off form factor is multiplied by the form-factor of the $\Delta$ resonance, as is suggested in Refs.~\cite{Luis_FF, cutoff}:
        \begin{eqnarray}
        F(p_\Delta)= \lambda^4_{\Delta}/(\lambda_{\Delta}^4 + (p^2_{\Delta} - M^2_{\Delta} )^2  ),
        \end {eqnarray}
        where $p_{\Delta}$ and $M_{\Delta}$ are the mass and the momentum of the $\Delta$ resonance and $\lambda_{\Delta}$ is a free parameter.
        The $\lambda_{\Delta}$ parameter and other adjusted parameters related to spin $3/2$ resonances are reported in \Cref{3half_fit}.
        \begin{table}
          \centering
          \caption{Fit parameters (spin-3/2 resonances). Dependent
            parameters are tabulated in the lower part of the table.}
          \label{3half_fit}
          \renewcommand{\arraystretch}{1.3}
          \begin{ruledtabular}
            \begin{tabular}{cccc|}
              Parameters & $P_{33}(1232)$&$D_{13}(1520)$
              \\ [0.1ex]
              \hline
              $a_{14}^{V}$ & -1.4533 & -1.410\\
              $ C^V_{3}(0)$ & 1.98 & 2.70\\
              $C^V_{4}(0)$ &-1.4285&-2.598 \\
              $C^V_{5}(0)$ &0.0&0.883 \\
              $\Lambda^V$ & 0.26&0.116\\
              $h_3^V$ & -0.0415&0.1751\\
              $k_3^V$ & 0.0216&-0.0338\\
              $h_4^V$ & -0.0642&0.0\\
              $k_4^V$ & 0.0341&0.0\\
              $h_5^V$ & ---&0.0\\
              $k_5^V$ & ---&0.0\\
              \hline
              $a_{11}^{V}$ & 2.0897 & 2.3544\\
              $a_{12}^{V}$ & -3.4146& -4.5682\\
              $a_{13}^{V}$ & 3.7782 & 4.6238\\
              $a_{21}^{V}$ & 2.101& 2.142\\
              $a_{22}^{V}$ & -3.586 & -3.41\\
              $a_{23}^{V}$ & 4.033 & 3.146\\
              $a_{24}^{V}$ & -1.548& -0.879\\
            \end{tabular}
          \end{ruledtabular}
        \end{table}
        \subsection{VMD form-factors for spin-$1/2$ resonances}
        There are two spin-$1/2$ resonances in the second resonance region, i.e., $P_{11}(1440)$ and $S_{11}(1535)$ with the form-factors in Eq.~(\ref{1half}). Five mesons in \Cref{meson} are taken into account, as suggested in Ref.~\cite{FF_QCD}. Therefore, $K=5$ for spin $1/2$ resonances. As a result there will be two free parameters in $b_{1 k}$ due to the three superconvergence relations [Eq.~(\ref{converg3})], and one free parameter will be allowed for $b_{2 k}$ due to the four superconvergence relations in Eq.~(\ref{converg4}). The QCD scale in the logarithmic renormalisations ($L^{(p)}_{\beta}$) can vary between $\Lambda_{QCD} \in [0.19-0.24]~\text{GeV}$ and $n_{\beta=1}=3$ and $n_{\beta=2}=2$.
        All adjusted parameters related to spin-$1/2$ resonances are reported in \Cref{1half_fit}. 
        \begin{table}
          \centering
          \caption{Fit parameters (spin-$1/2$ resonances). Dependent
            parameters are tabulated in the bottom part of the table.}
          \label{1half_fit}
          \renewcommand{\arraystretch}{1.3}
          \begin{ruledtabular}
            \begin{tabular}{cccc|}
              Parameters &  $P_{11}(1440)$&$S_{11}(1535)$
              \\ [0.1ex]
              \hline
              $b_{14}^{V}$ & 48.8104 & -35.7166\\
              $b_{15}^{V}$ & -15.7527 & 44.4834\\
              $b_{25}^{V}$ & 9.25904 & -23.5822\\
              $ g^V_{1}(0)$ & -0.0597& 49.7948\\
              $g^V_{2}(0)$ &44.5325 & 0.05281\\
              $\Lambda^V$ & 0.1172&0.116\\
              $h_1^V$ & -0.49584&25.5208\\
              $k_1^V$ & 0.06403&-1.8348\\
              $h_2^V$ & 7.3580&-0.7442\\
              $k_2^V$ & -0.5423&0.1412\\
              \hline
              $b_{11}^{V}$ & 3.1765 & -42.9822 \\
              $b_{12}^{V}$ & 10.6432 & 187.041\\
              $b_{13}^{V}$ & -45.8774 &-151.826 \\
              $b_{21}^{V}$ & 5.8018 & -7.1802\\
              $b_{22}^{V}$ & -34.9428 & 76.908\\
              $b_{23}^{V}$ & 61.6342 & -145.823\\
              $b_{24}^{V}$ & -40.7521 & 100.677\\
            \end{tabular}
          \end{ruledtabular}
        \end{table}
        \subsection{Dipole form-factors for resonances in the third region}
        The characteristic feature of the third region is overlapping structures of several resonances, each with a smaller overall contribution to the hadronic tensor.  
        The Rein-Sehgal model with a simple dipole form-factor for each resonance is used to avoid having a relatively large number of parameters (as is done with VMD form-factors) in a small phase space. The hadron current of each resonance in the third resonance region was related to a dipole form-factor with two adjustable parameters, $F_V(0)$ and $M_V$:
        \begin{eqnarray}
          F_V(k^2) &=& F_V(0)\left( 1- \frac{k^2}{M_V^2}\right)^{-2} \left(1 - \frac{k^2}{M^2}\right)^{n/2}\nonumber\\
        \end{eqnarray}	
        where $n$ is the number of oscillators from the Rein-Sehgal model. The number of oscillators and parameters $F_V(0)$ and $M_V$ are reported in \Cref{res_third} for resonances in the third region.
        \begin{table}
          \centering
          \caption{Fit result for the third resonance region   }
          \label{res_third}
          \renewcommand{\arraystretch}{1.3}
          \begin{ruledtabular}
            \begin{tabular}{lccccl}
              Resonance & $F^V(0)$ &$M_V(\text{GeV})$& $n$  
              \\ [0.1ex]
              \hline
              $P_{33}(1600)$ & 2.399 & 1.01&2\\
              $S_{31}(1620)$ & 0.10 &0.10&1 \\
              $F_{15}(1680)$&1.362& 0.692&2 \\
              $D_{33}(1700)$& 1.297 &0.10&1\\
              $  P_{11}(1710)$&0.10&0.10&2\\
              $  P_{13}(1720)$ & 3.00 &0.278&2\\
              $ F_{35} (1905)$ & 3.00 &1.141&2\\
              $ P_{31} (1910)$ & 3.00 &0.697&2 \\
              $ P_{33}(1920)$ &  0.10 &10.00&2 \\
              $ F_{37}(1950)$ & 3.00&0.576&2 \\
            \end{tabular}
          \end{ruledtabular}
        \end{table}
        \subsection{The VMD form-factors for non-resonant interactions}
        The update for non-resonant interaction was presented in sec.~\ref{nonres_VMD}.
        The adjustable parameters for non-resonant form-factors are defined in Eqs.~(\ref{GK1})-(\ref{GKFF2}) and the adjustable parameters for Regge propagators are defined in Eq.~(\ref{regge}). All parameters related to non-resonant interactions and their best fits are reported in \Cref{bkg_fit}.
        \begin{table}
          \centering
          \caption{Fit result for non-resonant interactions}
          \label{bkg_fit}
          \renewcommand{\arraystretch}{1.3}
          \begin{ruledtabular}
            \begin{tabular}{cc|ccl}
              Parameters & Best-fit value  &Parameters&  Best-fit value
              \\ [0.1ex]
              \hline
              $\mathcal{R}_{\rho'}$ & -1.358 & $N$& 1.20\\
              $\kappa{\rho'}$ & -3.821 & $\Lambda_{QCD}$& 0.260\\
              $\mathcal{R}_{\omega}$ & 0.862 & $\Lambda_D$& 2.111\\
              $\kappa_{\omega}$ & -49.899 & $\Lambda_1$& 0.841\\
              $\mathcal{R}_{\phi}$ & -7.605 & $\Lambda_2$& 0.419\\
              $\kappa_{\phi}$ & 0.081 & $S^{\pi}_0$& 3.995 \\
              $\mu_{\phi}$& 0.130 & $S^{\rho}_0$ & 0.769
            \end{tabular}
          \end{ruledtabular}
        \end{table}
        
        
        \section{Results and comparison with experimental data\label{result}}
        
        In this section the fit results from the updated MK model including the $1\sigma$ error band is presented. The systematic error band is constructed by randomly throwing the complete set of fit parameters within their post-fit correlated uncertainties (as encoded by the post-fit covariance matrix). The 1$\sigma$ error on each bin, and bib to bin correlations are calculated from the range containing $68\%$ of the throws closest to the best-fit model prediction. 
        The standard cross-section formula for the single pion electro-production in the resonance rest frame is the following: 
    \hspace*{0.cm}\vbox{\begin{align}
     &\frac{d^5 \sigma_{ep \rightarrow e' \pi N}}{dE_{e'} d\Omega_{e'} d\Omega^{\ast} _{\pi}}\nonumber\\
    & =\Gamma_{em} \Big[ \frac{d\sigma_T}{d\Omega^{\ast} _{\pi} }   + \epsilon \frac{d\sigma_L}{d\Omega^{\ast} _{\pi} }  + \sqrt{2\epsilon(1+ \epsilon)} \frac{d\sigma_{LT}}{d\Omega^{\ast} _{\pi} } \cos \phi^{\ast}_{\pi}\nonumber\\
  & + \epsilon \frac{d\sigma_{TT}}{d\Omega^{\ast} _{\pi} }\cos 2\phi^{\ast}_{\pi}  + h_e \sqrt{2\epsilon(1+ \epsilon)} \frac{d\sigma_{LT'}}{d\Omega^{\ast} _{\pi} } \sin \phi^{\ast}_{\pi}  \Big]~,
 \label{em_Xsec}
  \end{align}}  
    where $\Omega^{\ast} _{\pi}$ is the polar and azimuthal pion angles in the resonance rest frame and
  \begin{eqnarray}
  \Gamma_{em} = \frac{\alpha}{2\pi^2 Q^2} \frac{E_{e}}{E_{e'}} \frac{q_{\gamma}}{1-\epsilon} 
  \end{eqnarray}  
    is the virtual photon flux factor with $q_{\gamma}=(W^2 - M^2)/2M$, and $\epsilon = [1 + 2(q_{\gamma}^2 /Q^2) \tan^2 (\delta/2)]^{-1}$ where $\delta$ is the scattering angle of electron. \\
        
         Some SPP models were developed prior to the MK model. They used different theoretical calculations and parametrisation, but fit to the same experimental data. They include the following:
         \begin{itemize} 
        \item Unitary isobar model: MAID2007 \cite{MAID2007} is the latest version of unitary isobar model for partial wave analysis on the world data of pion photo and electro-production in the resonance region ($Q^2<5.0 ~ (\text{GeV}/c)^2$). 
        \item Dynamical coupled-channels (DCC) model \cite{DCC,DCC2013,DCC2016, ANL_Osaka, DCC_web}: DCC analysis uses the limited pion photo and electro-production data ($Q^2<3.0 ~ (\text{GeV}/c)^2$).
        \item Hybrid model \cite{hybrid}: Only valid at low $Q^2$, the Rarita-Schwinger formalism plus Regge approach is used for the resonant and non-resonant interactions like the MK model. However the form-factors in the two models are different. The form-factors in the Hybrid model are from a fit to the MAID result \cite{Olga_fit}.
        \end{itemize}
        \begin{figure*}
          \centering
          {\begin{minipage}{1.02\textwidth}
              \includegraphics[width=\textwidth]{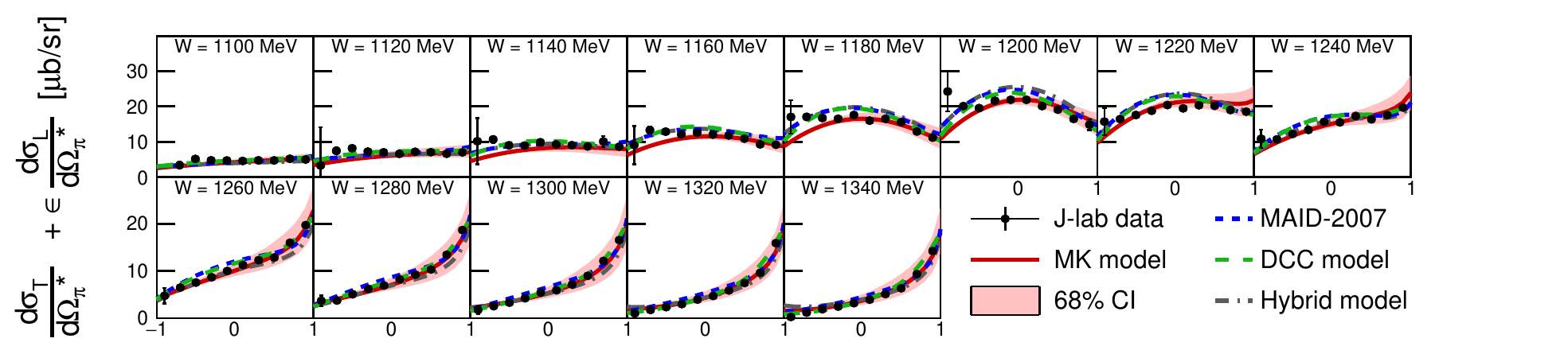}
                  \vspace*{0.5ex}
            \end{minipage}
            \begin{minipage}{1.02\textwidth}
              \includegraphics[width=\textwidth]{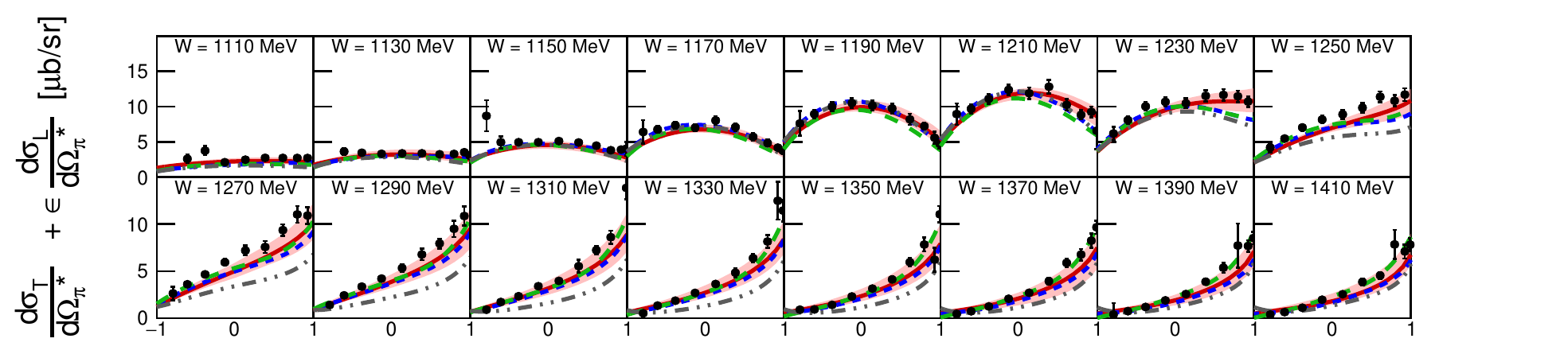}
                      \vspace*{0.5ex}
         \end{minipage}
            \begin{minipage}{1.02\textwidth}
              \includegraphics[width=\textwidth]{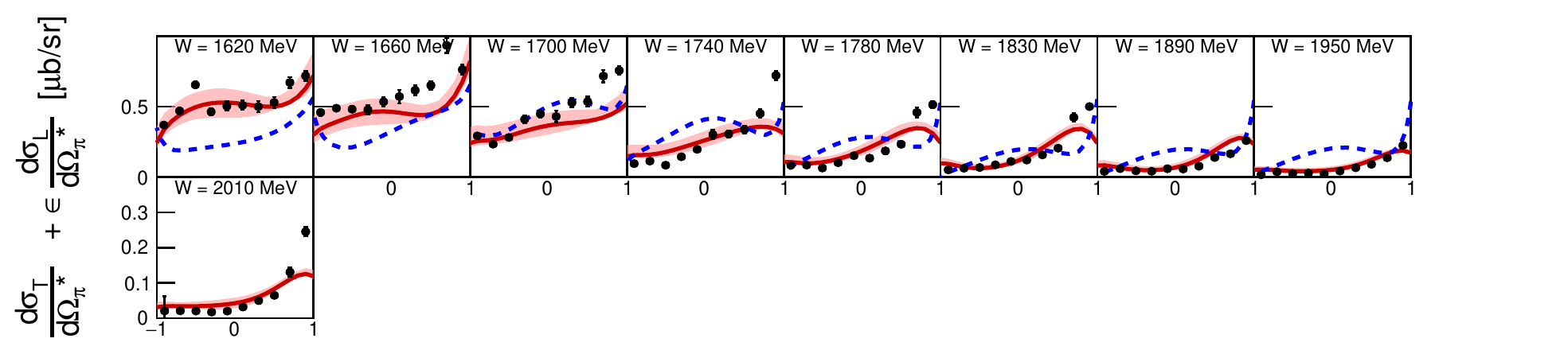}
             \vspace*{-4ex}
            \end{minipage}
            \begin{minipage}{1.02\textwidth}
              \includegraphics[width=\textwidth]{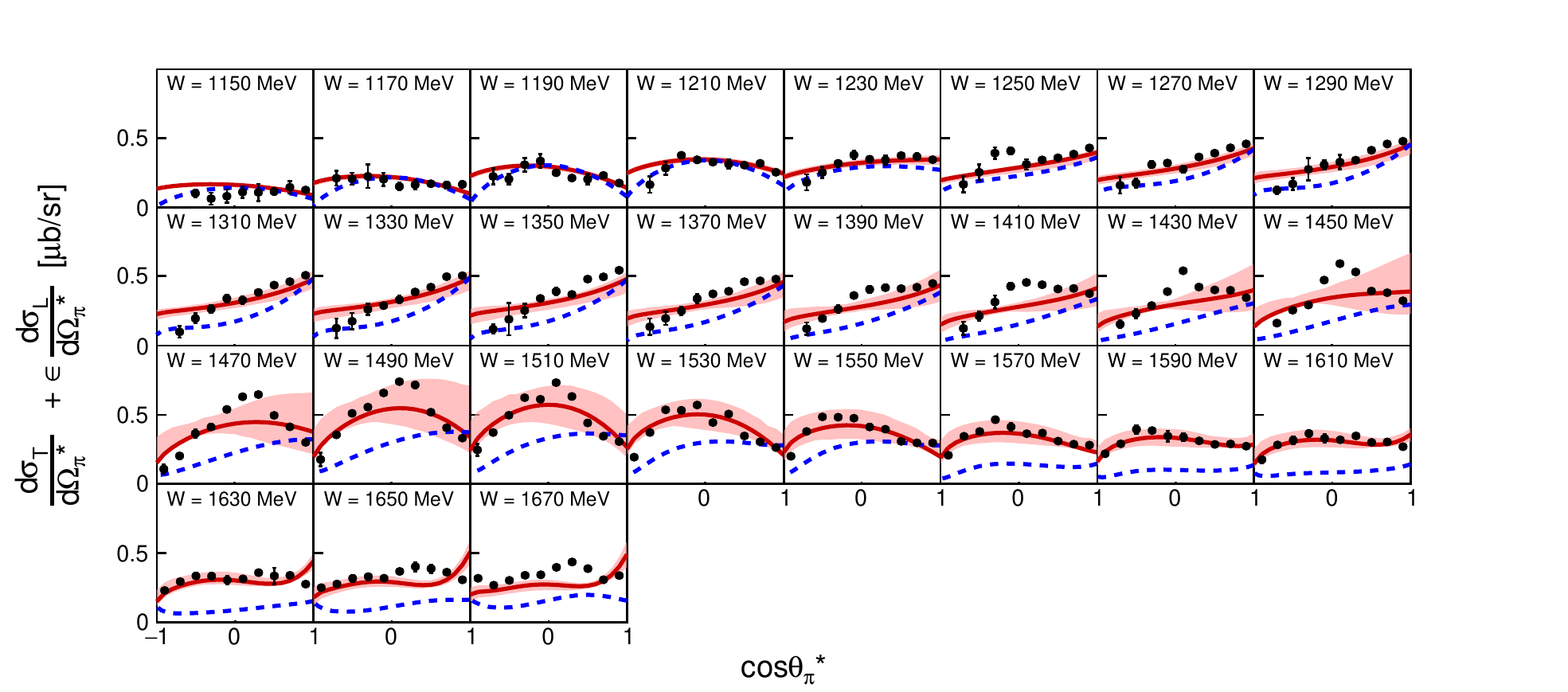}
            \end{minipage}
          }
          \vspace*{4ex}
          \caption{ Data-model comparisons for the $e p \rightarrow e n \pi^{+}$ channel at $Q^2= 0.2,~0.6,~ 2.6$ and $3.48 ~ (\text{GeV}/c)^2$, from top to bottom, respectively. The MK model result is the solid red line and the $68\%$ confidence interval is shown by the shaded band. The MAID, DCC and Hybrid results are shown by dashed blue, long-dashed green and dash-dotted gray lines respectively. } \label{npip_th}
        \end{figure*}
        \begin{figure*}
          \centering
          \vspace*{-2ex}
          {\begin{minipage}{1.02\textwidth}              \includegraphics[width=\textwidth]{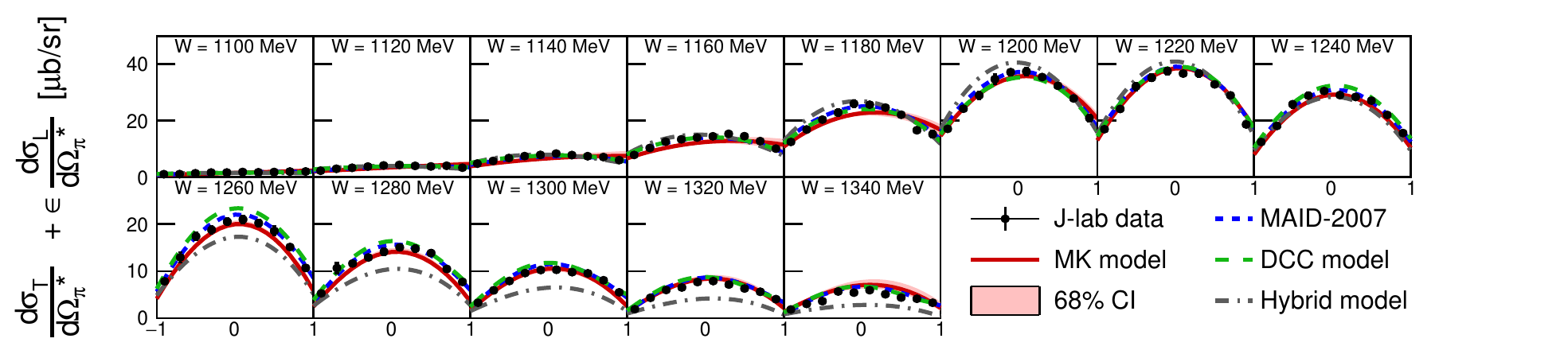}
              \vspace*{-4ex}
            \end{minipage}
            \begin{minipage}{1.02\textwidth}
              \includegraphics[width=\textwidth]{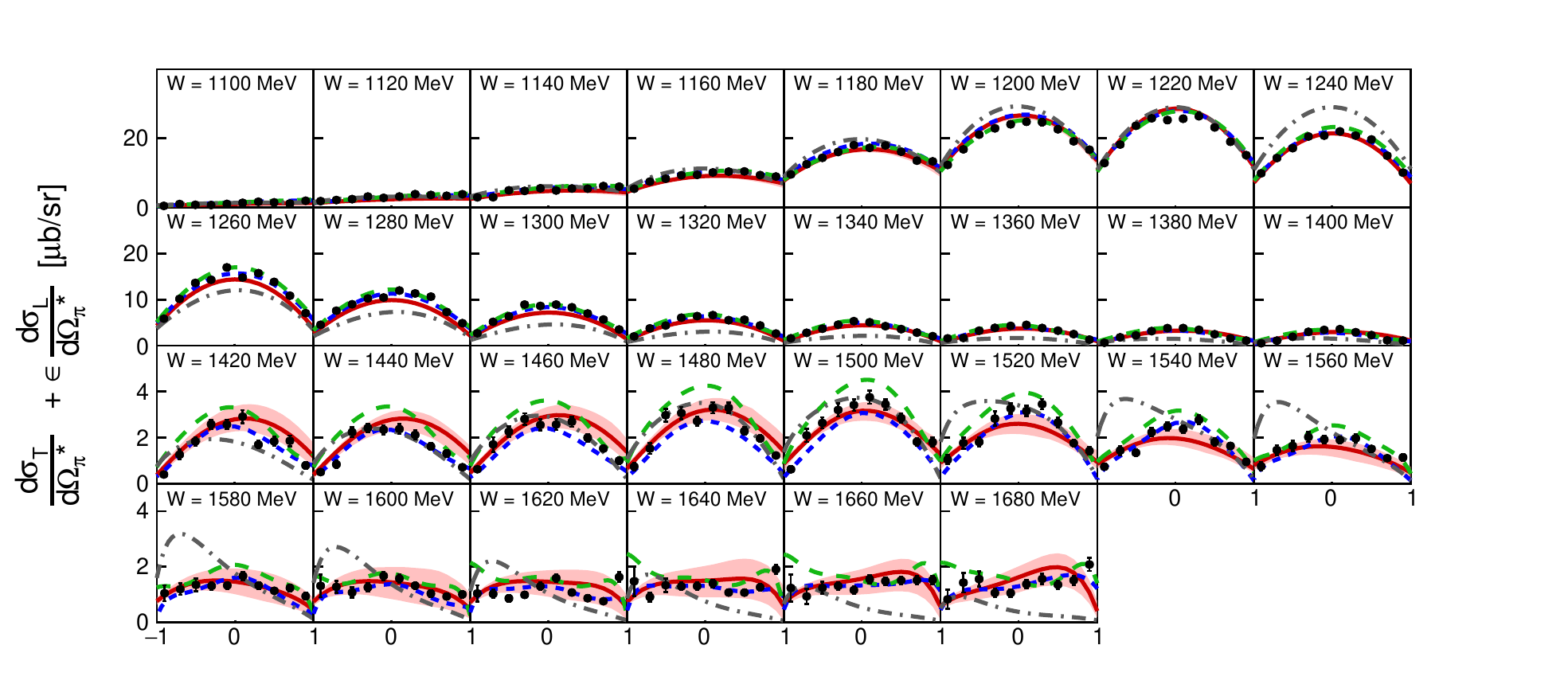}
                    \vspace*{-4ex}
         \end{minipage}
         \begin{minipage}{1.02\textwidth}
              \includegraphics[width=\textwidth]{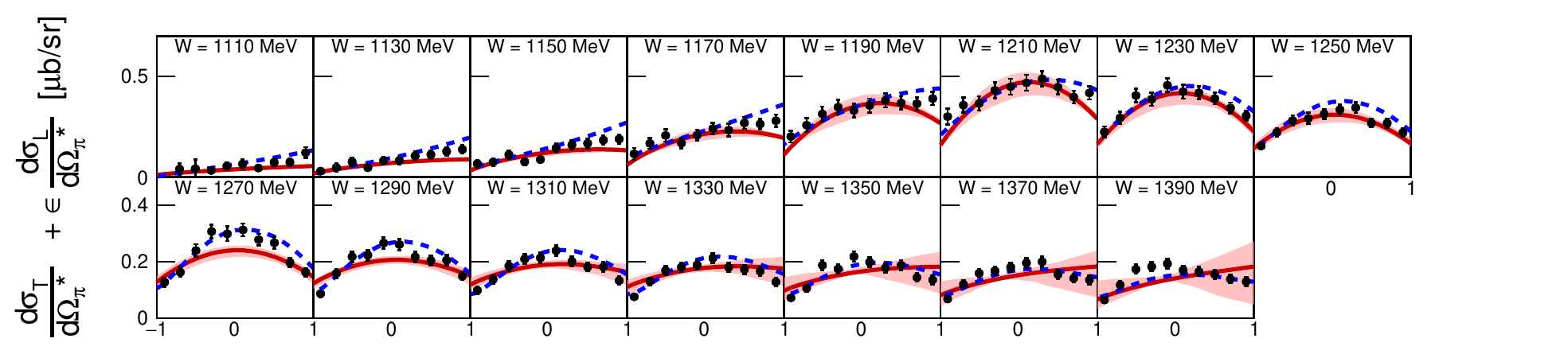}
            \end{minipage}
         \begin{minipage}{1.02\textwidth}
              \includegraphics[width=\textwidth]{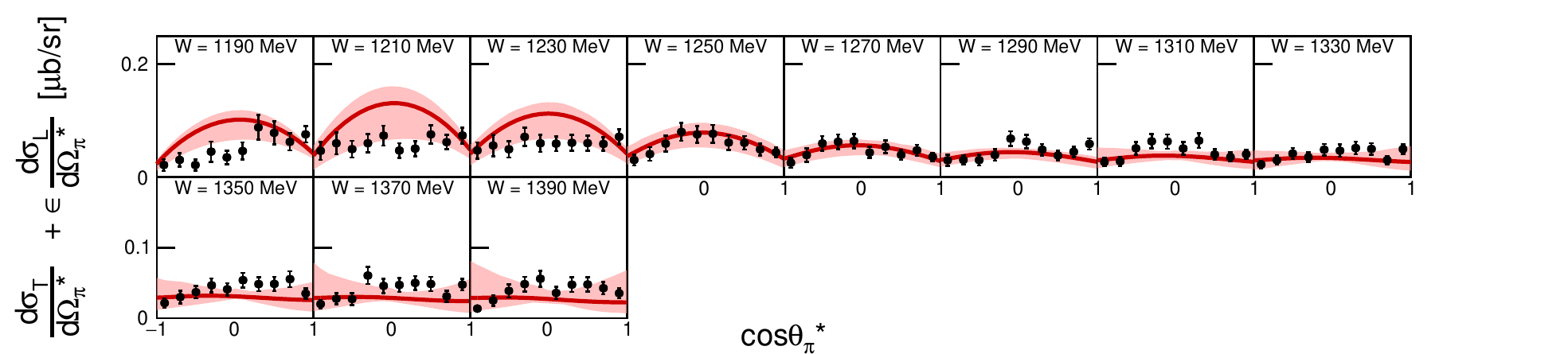}
          \end{minipage}
          }
          \vspace*{5ex}
          \caption{Data-model comparisons for the $e p \rightarrow e p \pi^{0}$ channel at $Q^2= 0.16,~0.4,~ 3.5$, and $6.00 ~ (\text{GeV}/c)^2$, from top to bottom, respectively. The MK model result is the solid red line and the $68\%$ confidence interval is shown by the shaded band. The MAID, DCC and Hybrid results are shown by the dashed blue, long-dashed green and dash-dotted gray lines respectively. } \label{ppi0_th}
        \end{figure*}

\begin{figure*}
          \centering
          {\begin{minipage}{1.09\textwidth}
                \includegraphics[width=\textwidth]{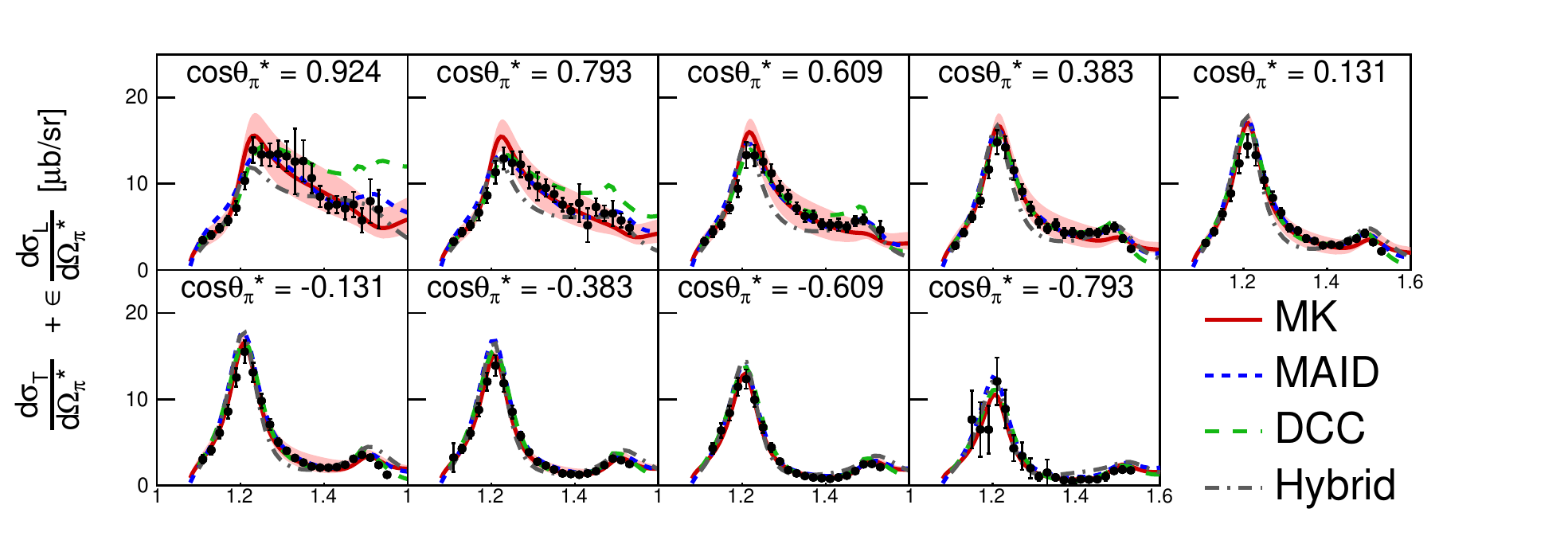}
                  \vspace*{-6ex}
            \end{minipage}
            \begin{minipage}{1.09\textwidth}
                 \includegraphics[width=\textwidth]{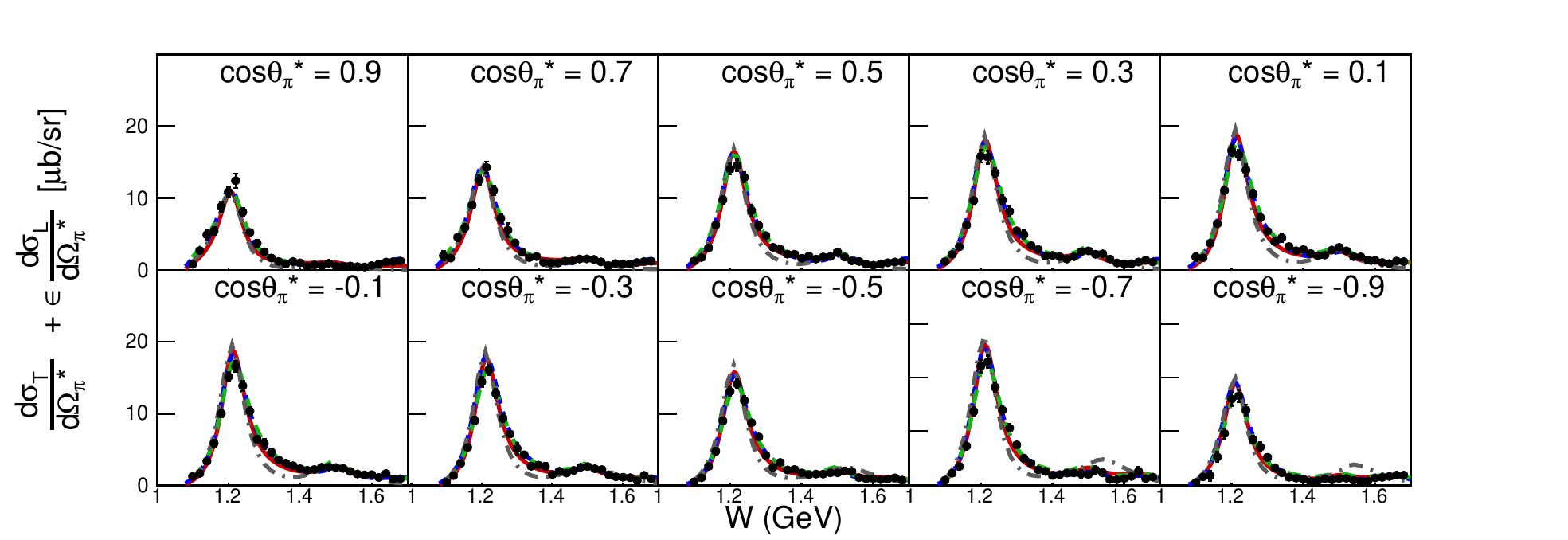}
            \end{minipage}}
          \caption{Data-model comparisons for low $Q^2$; $Q^2= 0.4 ~ (\text{GeV}/c)^2$ for the $e p \rightarrow e n\pi^{+}$ channel (top) and $Q^2= 0.65 ~ (\text{GeV}/c)^2$ for the $e p \rightarrow e p \pi^{0}$ channel (bottom). The MK model result is shown by the solid red line and the $68\%$ confidence interval is shown by the shaded band. The MAID, DCC and Hybrid results are shown by the dashed blue, long-dashed green, and dash-dotted gray lines respectively.
        } \label{npip_W}
          \vspace*{-0.1ex}
        \end{figure*}

\begin{figure*}
          \centering
          {\begin{minipage}{1.09\textwidth}
              \includegraphics[width=\textwidth]{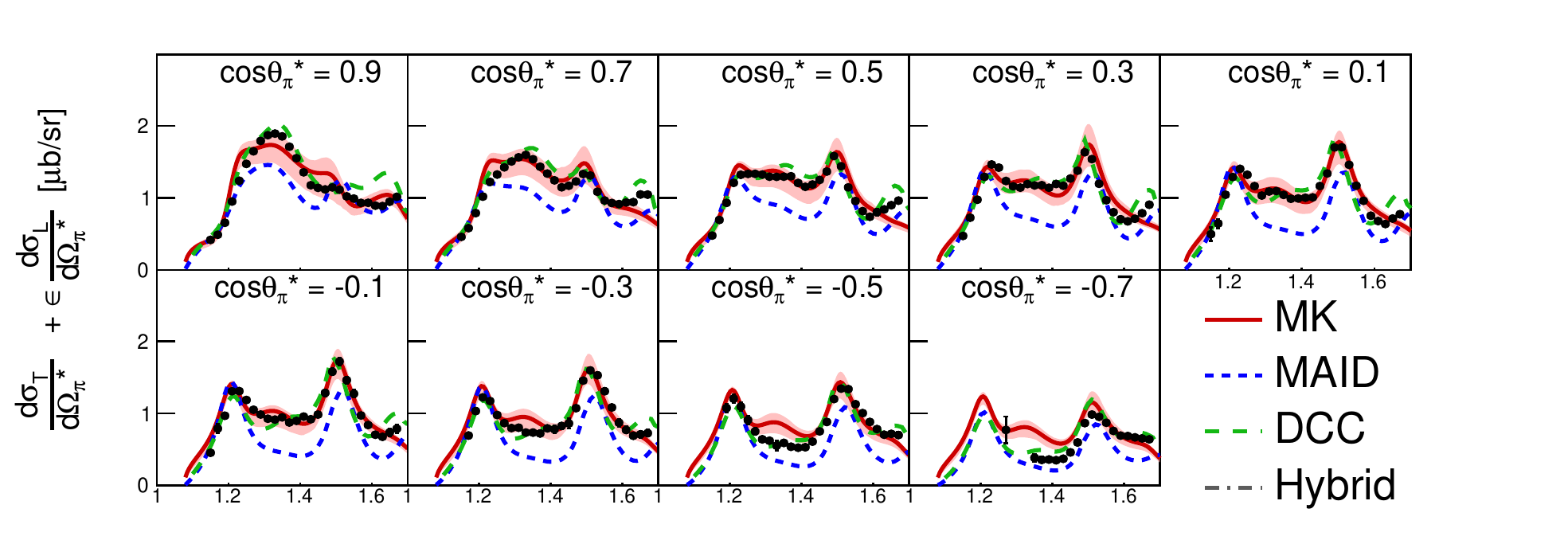}
                        \vspace*{-6ex}
            \end{minipage}
            \begin{minipage}{1.09\textwidth}
              \includegraphics[width=\textwidth]{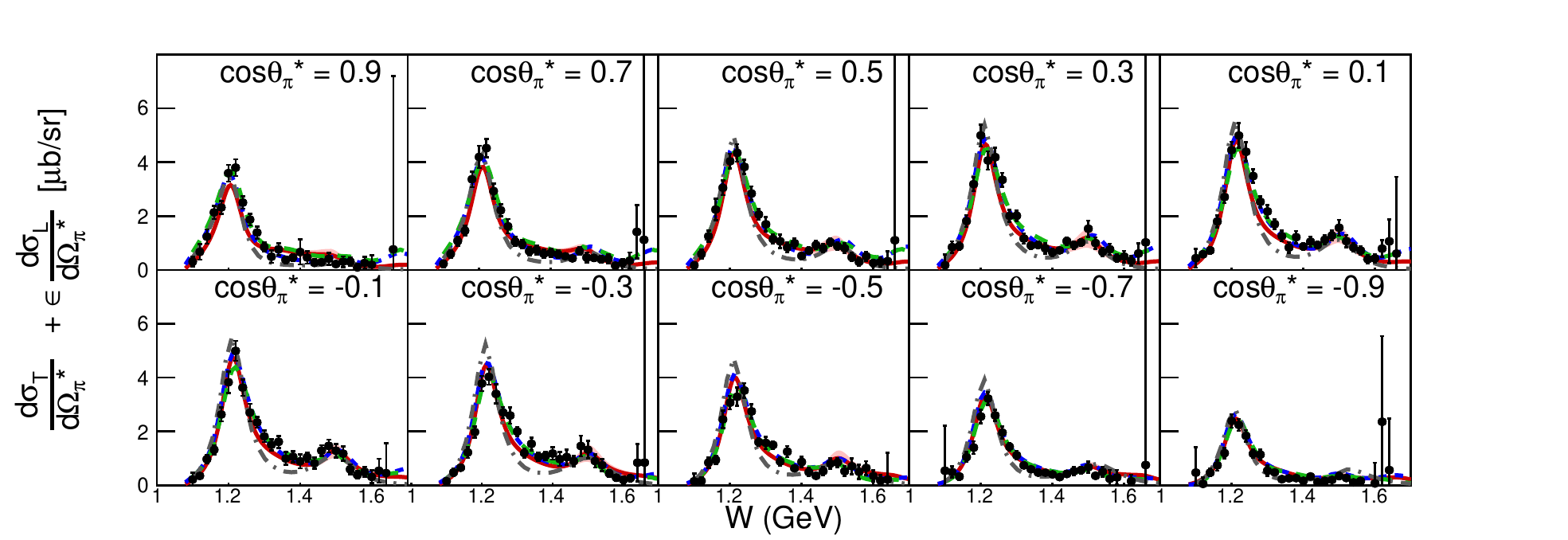}
            \end{minipage}}
         \caption{Data-model comparisons for medium $Q^2$; $Q^2= 2.05 ~ (\text{GeV}/c)^2$ for the $e p \rightarrow e pn\pi^{+}$ channel (top) and $Q^2= 1.45 ~ (\text{GeV}/c)^2$ for the $e p \rightarrow e p \pi^{0}$ channel (bottom). The MK model result is shown by the solid red line and the $68\%$ confidence interval is shown by the shaded band. The MAID, DCC and Hybrid results are shown by the dashed blue, long-dashed green and dash-dotted gray lines, respectively.} \label{ppi0_W}
          \vspace*{-0.1ex}
        \end{figure*}

        The selected results, in Figs.~\ref{npip_th} and~\ref{ppi0_th}, are chosen to show a broad range of $Q^2$ ($\in[0.16-6.00]~(\text{GeV}/c)^2$) and $W$ ($\in[1.1-2.01]~\text{GeV}$) for two channels with $p\pi^{0}$ and $n\pi^{+}$ final-state hadrons.
        Note that the valid region for the above models is smaller than that for the MK model. Therefore, the missing results from the above models in  Figs.~\ref{npip_th}~-~\ref{ppi0_W} are due to the model's stated limitations.\\
        
        \Cref{npip_th} shows the cross section for the $e p \rightarrow e n \pi^{+}$ channel and Fig.~\ref{ppi0_th} shows the cross section for the $e p \rightarrow e p \pi^{0}$ channel  in terms of the pion polar angle in the resonance rest frame, $\cos\theta^{\ast}$. Plots at very low invariant mass ($W<1.15~\text{GeV}$) are dominated by non-resonant background contributions where all models show relatively good agreement with data. At higher $W$, the $\Delta$ resonance has a dominant contribution in plots with $W \in [1.18- 1.3~\text{GeV}]$ where the Hybrid model deviates from the data above the $\Delta$ resonance's peak ($W>1.23~\text{GeV}$) as seen in the second group of plots in Figs.~\ref{npip_th} and \ref{ppi0_th}.\\
        
         At still higher invariant mass of the second resonance region ($W \in [1.4- 1.6~\text{GeV}]$), the next three resonances [$P_{11}(1440)$, $D_{13}(1520)$, and $S_{11}(1535)$] dominate. The second group of plots in Fig.~\ref{ppi0_th} shows that the DCC results overestimate the data but the shape agrees with the data. The Hybrid model doesn't predict the backward pion well. It also under-predicts the data in the dip region between the first and the second resonance regions, i.e., $W \in [1.28- 1.4~\text{GeV}]$, which is visible in Figs.~\ref{npip_th} and \ref{ppi0_th}.\\
         
         The MAID results \cite{MAID_web} show agreement with data at low $Q^2$ (two top groups of plots in Fig.~\Cref{npip_th,ppi0_th}); however, at higher $Q^2$, the result favours data from the $e p \rightarrow e p \pi^{0}$ channel over the $e p \rightarrow e n \pi^{+}$ channel. \\
         
         The rest of the resonances in the \Cref{res_list} contribute to the third resonance region ($W > 1.6~\text{GeV}$) and only one measurement for the $e p \rightarrow e n \pi^{+}$ channel exists. This is presented in the third group of plots in Fig.~\ref{npip_th} where the MAID results deviate by more than $3\sigma$ from the data.\\
         
        Some of the datasets in \Cref{clas_data} are over a large range of hadron invariant mass  which will allow the cross section in different resonance regions to be shown. See Fig.~\ref{npip_W} (Fig.~\ref{ppi0_W}) for low (medium) $Q^2$. Comparing the two channels in each figure reveals the different $W$ distributions, which are mainly due to the different Clebsch-Gordan coefficients for isospin-$3/2$ and -$1/2$ resonances. The first resonance region is dominant in the $e p \rightarrow e p \pi^{0}$ channel due to larger Clebsch-Gordan coefficients for the $\Delta$ (isospin-$3/2$) resonance. The second resonance region is more pronounced for the $e p \rightarrow e n \pi^{+}$ channel because this region is populated mainly by isospin-$1/2$ resonances that have larger Clebsch-Gordan coefficients for this channel. The isospin Clebsch-Gordan coefficients are given in \Cref{CG}.\\
        
        All comments about the model comparisons in the first and the second resonance regions from Figs.~\ref{npip_th} and \ref{ppi0_th} are visible in Figs.~\ref{npip_W} and \ref{ppi0_W}. The third resonance region is not visible here. The DCC prediction for the $e p \rightarrow e n \pi^{+}$ channel at low $Q^2$ and forward pion angles (top plots in Fig.~\ref{npip_W}), overestimate the data in the second resonance region.  
        \section{conclusion}
        In this work the vector current description of the MK model is improved by extending the valid region of the model by using all exclusive single pion electo-production data and advanced analysis techniques. The MK model is extended to high momentum transfer by using the vector meson dominance form-factor model consistent with the quantum chromodynamics theory through quark-hadron duality. It is also extended to high hadron invariant mass by using Regge phenomenology, i.e., by replacing the \emph{t}-channel Feynman propagators with the corresponding Regge trajectories. \\
        
        The free parameters of the model are constructed by fitting to exclusive electron scattering data in a large kinematic region. The free parameters determine the behaviours of the vector form factors of resonances and the proton in the non-resonant interaction. An advanced analysis method with the improved parametrisation  produces a systematic uncertainty tied to data and minimises bias. \\
        
        The goodness-of-fit and the fit results of the analysis used in defining the MK model (Figs.~\ref{npip_th} - \ref{ppi0_W}) in a broad range of $Q^{2}$ ($\in[0.16-6.00]~(\text{GeV}/c)^2$) and $W$ ($\in[1.1-2.01]~\text{GeV}$) show  the resulting nucleon form factors are adequate to describe the data for both channels. The overall agreement with data demonstrates that the VMD model can describe form factors in both perturbative and non-perturbative domains. Therefore, reasonable behaviour of the MK model now extends outside the resonant region and into the transition region between the resonance and DIS regions. The results show some disagreements with the data, especially in regions with high $Q^2$ (see the bottom plots of Fig.~\ref{ppi0_th}), but the error bands generally accommodate the disagreements. This feature is of crucial importance to experimental efforts in need of a SPP model.\\
        
        Other models were also compared with data in their own valid regions. At low $Q^2$, the data (from both channels) are well described by the MAID model. At higher $Q^2$, the MAID results show the data agreement for the $e p \rightarrow e p \pi^{0}$ channel is better than that for the $e p \rightarrow e n \pi^{+}$ channel. Since in theory the only difference between the two channels is the Clebsch-Gordan coefficients, this demonstrates a bias in their analysis.\\
        
        The DCC results overestimate data in the second resonance region, but the deviation is small as can be seen in the second group of plots in Fig.~\ref{ppi0_th}. However, the model agrees with data in its stated valid region [$Q^2<3.0 ~ (\text{GeV}/c)^2$]. The Hybrid model predictions, which are not a direct fit result to the experimental data, are more than $3\sigma$ away from the experimental data, mainly in the second resonance and transition regions.  \\
        
        The hadron currents for resonant and non-resonant interactions from this work will be used in the hadron vector current of the neutrino-nucleon cross-sections. The electron scattering data used in this work covers the kinematic regions essential for all accelerator-based long-baseline neutrino experiments.

        \begin{acknowledgments}
          I am extremely grateful to Deborah Harris and Dave Wark  for being a constant source of encouragement and support. I appreciate the support from my colleagues at Fermilab. I would like to thank the York neutrino group, Stephen Dolan, and Daniel Cherdack for fruitful discussions. I am also thankful to L. Cole Smith, Toru Sato, and Alexis Nikolakopoulos for providing me with data and the model's predictions for single pion electro-production. This work was supported by the NSERC and the Royal Commission for the Exhibition of 1851.
        \end{acknowledgments}
        \appendix
        \section{Helicity amplitudes for the resonant interactions}\label{appA}
        The helicity amplitudes of the vector current are given in \Cref{res_HV}, where $f^V(R)$ is the production amplitude given in Eqs. (\ref{delta})-(\ref{last}) and $D(R)$ is the decay amplitude:
        \begin{table*}
        \centering
        \caption{Vector helicity amplitudes of resonant interaction. }
        \label{res_HV}
        \renewcommand{\arraystretch}{1.3}
        \begin{ruledtabular}
         \begin{tabular}{c|c|c|c}
          $\lambda_2$& $\lambda_1$ &$\tilde{F}_{\lambda_2 \lambda_1}^{e_{L}}(\theta, \phi)$& $\tilde{F}_{\lambda_2 \lambda_1}^{e_{R}}(\theta, \phi)$\\[4pt]
         \hline
         $\begin{aligned}
        &\scalebox{0.001}{~}\\&\frac{1}{2}\\[7pt] -&\frac{1}{2} \\[7pt] &\frac{1}{2}\\[7pt] -&\frac{1}{2}
        \end{aligned}$
        &
         $\begin{aligned}
        &\scalebox{0.001}{~}\\&\frac{1}{2}\\[7pt] &\frac{1}{2} \\[7pt] -&\frac{1}{2}\\[7pt] -&\frac{1}{2}
        \end{aligned}$
        &
         $\begin{aligned}
        &\scalebox{0.001}{~}\\
        \pm&\sum_j\frac{2j+1}{\sqrt{2}} \mathcal{D}^j(R) ~f^V_{-3}(R(I,j=l\pm\frac{1}{2}))~  d^j_{\frac{3}{2} \frac{1}{2}}(\theta) e^{-2i\phi}\\
        &\sum_j\frac{2j+1}{\sqrt{2}} \mathcal{D}^j(R)~ f^V_{-3}(R(I,j=l\pm\frac{1}{2}))~  d^j_{\frac{3}{2} -\frac{1}{2}}(\theta)e^{-i\phi}\\
        \mp&\sum_j\frac{2j+1}{\sqrt{2}} \mathcal{D}^j(R)~ f^V_{-1}(R(I,j=l\pm\frac{1}{2}))~  d^j_{\frac{1}{2} \frac{1}{2}}(\theta)e^{-i\phi}\\
        -&\sum_j\frac{2j+1}{\sqrt{2}} \mathcal{D}^j(R)~ f^V_{-1}(R(I,j=l\pm\frac{1}{2}))~  d^j_{\frac{1}{2} -\frac{1}{2}}(\theta)\\
        \end{aligned}$
        &
         $\begin{aligned}
         &\scalebox{0.01}{~}\\
        -&\sum_j\frac{2j+1}{\sqrt{2}} \mathcal{D}^j(R)~ f^V_{-1}(R(I,j=l\pm\frac{1}{2}))~  d^j_{\frac{1}{2} -\frac{1}{2}}(\theta)\\
        \pm&\sum_j\frac{2j+1}{\sqrt{2}} \mathcal{D}^j(R)~ f^V_{-1}(R(I,j=l\pm\frac{1}{2}))~  d^j_{\frac{1}{2} \frac{1}{2}}(\theta)e^{i\phi}\\
        -&\sum_j\frac{2j+1}{\sqrt{2}} \mathcal{D}^j(R)~ f^V_{-3}(R(I,j=l\pm\frac{1}{2}))~  d^j_{\frac{3}{2} -\frac{1}{2}}(\theta)e^{i\phi}\\
        \pm&\sum_j\frac{2j+1}{\sqrt{2}} \mathcal{D}^j(R) ~f^V_{-3}(R(I,j=l\pm\frac{1}{2}))~  d^j_{\frac{3}{2} \frac{1}{2}}(\theta) e^{2i\phi}
        \end{aligned}$
        \\\hline
        &  &$\tilde{F}_{\lambda_2 \lambda_1}^{e_{-}}(\theta, \phi)$& $\tilde{F}_{\lambda_2 \lambda_1}^{e_{+}}(\theta, \phi)$\\[4pt]
         \hline
         $\begin{aligned}
        &\scalebox{0.001}{~}\\&\frac{1}{2}\\[7pt] -&\frac{1}{2} \\[7pt] &\frac{1}{2}\\[7pt] -&\frac{1}{2}
        \end{aligned}$
        &
         $\begin{aligned}
        &\scalebox{0.001}{~}\\&\frac{1}{2}\\[7pt] &\frac{1}{2} \\[7pt] -&\frac{1}{2}\\[7pt] -&\frac{1}{2}
        \end{aligned}$
        &
         $\begin{aligned}
        &\scalebox{0.05}{~}\\
        &\frac{|\mathbf k|}{\sqrt{-k^2}}\sum_j\frac{2j+1}{\sqrt{2}} \mathcal{D}^j(R)~ f^{V(-)}_{0+}(R(I,j=l\pm\frac{1}{2}))~  d^j_{\frac{1}{2} \frac{1}{2}}(\theta)e^{-i\phi}\\
        \pm&\frac{|\mathbf k|}{\sqrt{-k^2}}\sum_j\frac{2j+1}{\sqrt{2}} \mathcal{D}^j(R)~ f^{V(-)}_{0+}(R(I,j=l\pm\frac{1}{2}))~  d^j_{\frac{1}{2} -\frac{1}{2}}(\theta)\\
        \pm&\frac{|\mathbf k|}{\sqrt{-k^2}}\sum_j\frac{2j+1}{\sqrt{2}} \mathcal{D}^j(R)~ f^{V(-)}_{0+}(R(I,j=l\pm\frac{1}{2}))~  d^j_{\frac{1}{2} -\frac{1}{2}}(\theta)\\
        -&\frac{|\mathbf k|}{\sqrt{-k^2}}\sum_j\frac{2j+1}{\sqrt{2}} \mathcal{D}^j(R)~ f^{V(-)}_{0+}(R(I,j=l\pm\frac{1}{2}))~  d^j_{\frac{1}{2} \frac{1}{2}}(\theta)e^{i\phi}\end{aligned}$
        &
         $\begin{aligned}
         &\scalebox{0.05}{~}\\
        &\frac{|\mathbf k|}{\sqrt{-k^2}}\sum_j\frac{2j+1}{\sqrt{2}} \mathcal{D}^j(R)~ f^{V(+)}_{0+}(R(I,j=l\pm\frac{1}{2}))~  d^j_{\frac{1}{2} \frac{1}{2}}(\theta)e^{-i\phi}\\
        \pm&\frac{|\mathbf k|}{\sqrt{-k^2}}\sum_j\frac{2j+1}{\sqrt{2}} \mathcal{D}^j(R)~ f^{V(+)}_{0+}(R(I,j=l\pm\frac{1}{2}))~  d^j_{\frac{1}{2} -\frac{1}{2}}(\theta)\\
        \pm&\frac{|\mathbf k|}{\sqrt{-k^2}}\sum_j\frac{2j+1}{\sqrt{2}} \mathcal{D}^j(R)~ f^{V(+)}_{0+}(R(I,j=l\pm\frac{1}{2}))~  d^j_{\frac{1}{2} -\frac{1}{2}}(\theta)\\
        -&\frac{|\mathbf k|}{\sqrt{-k^2}}\sum_j\frac{2j+1}{\sqrt{2}} \mathcal{D}^j(R)~ f^{V(+)}_{0+}(R(I,j=l\pm\frac{1}{2}))~  d^j_{\frac{1}{2} \frac{1}{2}}(\theta)e^{i\phi}\end{aligned}$
        \\
        \end{tabular}
        \end{ruledtabular}
        \end{table*}
        \begin{align}\label{D_amp}
        \mathcal{D}^j(R)=\langle N\pi,\lambda_2|R \lambda_R \rangle &= \sigma^D C_{N\pi}^{j} \sqrt{\chi_E} \kappa C_{N\pi}^{I} f_{BW}~. 
        \end{align}
        where $f_{BW}(R)$ is the Breit-Wigner amplitude,
        \begin{eqnarray}
        f_{BW}(R) = \sqrt{\frac{\Gamma_R}{2\pi} }\left( \frac{1}{W- M_R + i\Gamma_R/2} \right )~,
        \end{eqnarray}
        where \begin{eqnarray}
        \Gamma_R = \Gamma_0 (|\mathbf{q}(W)|/|\mathbf{q}(M_R)|)^{2l+1},
        \end{eqnarray}
        and 
        \begin{equation}
        \kappa= \left( 2\pi^2 \frac{W^2}{M^2} ~.~\frac{2}{2j+1} ~ \frac{1}{|\mathbf{q}|} \right)^{\frac{1}{2}}.
        \end{equation}
        $\Gamma_0$, $M_R$, $\sigma^D$, and $\chi_E$ are given in \Cref{res_list} and $C^I_{N\pi}$ are the isospin Clebsch-Gordan coefficients given in \Cref{CG}.
        \begin{table}
        \centering
        \caption{Isospin Clebsch-Gordan coefficients} \label{CG}
        \renewcommand{\arraystretch}{1.5}
        \begin{ruledtabular}
         \begin{tabular}{|ccc|}
         {Channels} & $C^{3/2}_{N\pi}$& $C^{1/2}_{N\pi}$   \\ [0.7ex]
         \hline
        $e p \rightarrow e p \pi^{0}$    &  $\sqrt{\frac{2}{3}}$  &  $-\sqrt{\frac{1}{3}}$\\
        $e p \rightarrow e n \pi^{+}$   &$-\sqrt{\frac{1}{3}}$  &$-\sqrt{\frac{2}{3}}$\\
        \end{tabular}
        \end{ruledtabular}
        \end{table}
        The explicit forms of the $d^j_{\lambda,\mu}(\theta)$ functions for $j=l+\frac{1}{2}$ are:
        \begin{align}
        d^j_{\frac{1}{2} \frac{1}{2}}~&= (l+1)^{-1} \cos\frac{\theta}{2} (P'_{l+1} - P'_l)\nonumber\\
        d^j_{-\frac{1}{2} \frac{1}{2}}&= (l+1)^{-1} \sin\frac{\theta}{2} (P'_{l+1} + P'_l)\nonumber\\
        d^j_{\frac{1}{2} \frac{3}{2}}~&= (l+1)^{-1} \sin\frac{\theta}{2} (\sqrt{\frac{l}{l+2}}P'_{l+1} + \sqrt{\frac{l+2}{l}} P'_l)\nonumber\\
        d^j_{-\frac{1}{2} \frac{3}{2}}&= (l+1)^{-1} \cos\frac{\theta}{2} (-\sqrt{\frac{l}{l+2}}P'_{l+1} + \sqrt{\frac{l+2}{l}} P'_l)\nonumber
        \label{dj_def}
        \end{align}
        where $P_l$ are the Legendre polynomials and $P'_l= dP_l/d\cos\theta$.
        \section{Helicity amplitudes for the non-resonant interactions}\label{appB}
        \begin{table*}
        \centering
        \caption{Vector helicity amplitudes of non-resonant interaction. }
        \label{nonres_HV}
        \renewcommand{\arraystretch}{1.3}
        \begin{ruledtabular}
         \begin{tabular}{c|c|c|c}
          $\lambda_2$& $\lambda_1$ &$\tilde{F}_{\lambda_2 \lambda_1}^{e_{L}}(\theta, \phi)$& $\tilde{F}_{\lambda_2 \lambda_1}^{e_{R}}(\theta, \phi)$\\[4pt]
         \hline
         $\begin{aligned}
        &\scalebox{0.001}{~}\\&\frac{1}{2}\\[7pt] -&\frac{1}{2} \\[7pt] &\frac{1}{2}\\[7pt] -&\frac{1}{2}
        \end{aligned}$
        &
         $\begin{aligned}
        &\scalebox{0.001}{~}\\&\frac{1}{2}\\[7pt] &\frac{1}{2} \\[7pt] -&\frac{1}{2}\\[7pt] -&\frac{1}{2}
        \end{aligned}$
        &
         $\begin{aligned}
        &\scalebox{0.001}{~}\\
        &\frac{1}{\sqrt{2}} e^{-2i\phi} \sin{\theta} \cos\frac{\theta}{2} (\mathscr{F}_3 + \mathscr{F}_4)\\
        -&\frac{1}{\sqrt{2}} e^{-i\phi} \sin{\theta} \sin\frac{\theta}{2} (\mathscr{F}_3 - \mathscr{F}_4)\\
        &\sqrt{2} e^{-i\phi} \big[\cos\frac{\theta}{2} (\mathscr{F}_1 - \mathscr{F}_2) - \frac{1}{2} \sin{\theta} \sin\frac{\theta}{2} (\mathscr{F}_3 - \mathscr{F}_4)\big]\\
        -&\sqrt{2}  \big[\sin\frac{\theta}{2} (\mathscr{F}_1 + \mathscr{F}_2) + \frac{1}{2} \sin{\theta} \cos\frac{\theta}{2} (\mathscr{F}_3 + \mathscr{F}_4)\big]
        \end{aligned}$
        &
         $\begin{aligned}
         &\scalebox{0.01}{~}\\
         -&\sqrt{2}  \big[\sin\frac{\theta}{2} (\mathscr{F}_1 + \mathscr{F}_2) + \frac{1}{2} \sin{\theta} \cos\frac{\theta}{2} (\mathscr{F}_3 + \mathscr{F}_4)\big]\\
         -&\sqrt{2} e^{i\phi} \big[\cos\frac{\theta}{2} (\mathscr{F}_1 - \mathscr{F}_2) - \frac{1}{2} \sin{\theta} \sin\frac{\theta}{2} (\mathscr{F}_3 - \mathscr{F}_4)\big]\\
        &\frac{1}{\sqrt{2}} e^{i\phi} \sin{\theta} \sin\frac{\theta}{2} (\mathscr{F}_3 - \mathscr{F}_4)\\
         &\frac{1}{\sqrt{2}} e^{2i\phi} \sin{\theta} \cos\frac{\theta}{2} (\mathscr{F}_3 + \mathscr{F}_4)
        \end{aligned}$
        \\\hline
        &  &$\tilde{F}_{\lambda_2 \lambda_1}^{e_{-}}(\theta, \phi)$& $\tilde{F}_{\lambda_2 \lambda_1}^{e_{+}}(\theta, \phi)$\\[4pt]
         \hline
         $\begin{aligned}
        &\scalebox{0.001}{~}\\&\frac{1}{2}\\[7pt] -&\frac{1}{2} \\[7pt] &\frac{1}{2}\\[7pt] -&\frac{1}{2}
        \end{aligned}$
        &
         $\begin{aligned}
        &\scalebox{0.001}{~}\\&\frac{1}{2}\\[7pt] &\frac{1}{2} \\[7pt] -&\frac{1}{2}\\[7pt] -&\frac{1}{2}
        \end{aligned}$
        &
         $\begin{aligned}
        &\scalebox{0.05}{~}\\
        & e^{-i\phi}\cos\frac{\theta}{2}\frac{1}{C_{-}}(k_0\epsilon^0_L - |\mathbf{k}| \epsilon^3_L )(\mathscr{F}_5 + \mathscr{F}_6) \\
          -&\sin\frac{\theta}{2}\frac{1}{C_-}(k_0\epsilon^0_L - |\mathbf{k}| \epsilon^3_L )(\mathscr{F}_5 - \mathscr{F}_6) \\
         -&\sin\frac{\theta}{2}\frac{1}{C_-}(k_0\epsilon^0_L - |\mathbf{k}| \epsilon^3_L)(\mathscr{F}_5 - \mathscr{F}_6) \\
         -& e^{i\phi}\cos\frac{\theta}{2}\frac{1}{C_-}(k_0\epsilon^0_L - |\mathbf{k}| \epsilon^3_L)(\mathscr{F}_5 + \mathscr{F}_6)
        \end{aligned}$
        &
         $\begin{aligned}
         &\scalebox{0.05}{~}\\
        & e^{-i\phi}\cos\frac{\theta}{2}\frac{1}{C_{+}}(k_0\epsilon^0_R - |\mathbf{k}| \epsilon^3_R)(\mathscr{F}_5 + \mathscr{F}_6)\\
          -&\sin\frac{\theta}{2}\frac{1}{C_+}(k_0\epsilon^0_R - |\mathbf{k}| \epsilon^3_R)(\mathscr{F}_5 - \mathscr{F}_6)\\
         -&\sin\frac{\theta}{2}\frac{1}{C_+}(k_0\epsilon^0_R - |\mathbf{k}| \epsilon^3_R)(\mathscr{F}_5 - \mathscr{F}_6) \\
        - &e^{i\phi}\cos\frac{\theta}{2}\frac{1}{C_+}(k_0\epsilon^0_R - |\mathbf{k}| \epsilon^3_R)(\mathscr{F}_5 + \mathscr{F}_6)
        \end{aligned}$\\
        \end{tabular}
        \end{ruledtabular}
        \end{table*}
        The helicity amplitudes of the vector current are given in \Cref{nonres_HV}, where 
        \begin{eqnarray}
        \mathscr{F}_i = K^V_i F_i ~(i=1, ...,6),
        \end{eqnarray}
        where
        \begin{eqnarray}
        \begin{aligned}
        F_1&= V_1 + (V_3-V_4)(qk)/W_- +  V_4W_{-} - V_6 k^2/W_-~,\\
        F_2&=-V_1 + (V_3-V_4)(qk)/W_+ +  V_4W_{+} - V_6 k^2/W_+~,\\
        F_3 &= V_3 - V_4 + V_{25}/W_+ ~, \\
        F_4 &= V_3 - V_4 - V_{25}/W_- ~, \\
        F_5 &= V_1(W_+^2 - k^2)/2W - V_2(qk)(W_+^2 - k^2 + 2WW_-)/2W \\&+ (V_3-V_4)(W_+q_0 - (qk)) + V_4(W_+^2 - k^2)W_-/2W \\&- V_5(qk)k_0 - V_6 (W_{+}^2 - k^2)W_{-}/2W + q_0 V_{25}~,\nonumber
        \end{aligned}
        \end{eqnarray}
        \\
     \begin{eqnarray}
                F_6 &=-V_1(W_-^2 - k^2)/2W + V_2(qk)(W_+^2 - k^2 + 2WW_-)/2W \nonumber\\&+ (V_3-V_4)(W_-q_0 - (qk))  + V_4(W_-^2 - k^2)W_+/2W \nonumber\\&+ V_5(qk)k_0 - V_6 (W_{-}^2 - k^2)W_{+}/2W - q_0 V_{25}~,
                \nonumber\\&~
        \end{eqnarray}
        and $V_i$ $(i=1, ...,6)$ are presented in \Cref{inv_amp}, where $s,~u$, and $t$ are invariant Mandelstam variables:
        \begin{eqnarray}
        s = (p_2 + q)^2 = (p_1 + k)^2 = W^2~,\nonumber\\
        t = (k-q)^2~, ~~ \text{and}~~ u=(q-p_1)^2~.
        \end{eqnarray}
         $K^V_i$ are given in Ref. \cite{Adler}:
        \begin{equation}
        \begin{tabular}{ll}
        $\begin{aligned}
        K_1^V &= W_- O_{1+}\\
        K_2^V &= W_+ O_{1-}\\
        K_3^V &= q^2 W_+ O_{2-}
        \end{aligned}$
        &~~
        $\begin{aligned}
        K_4^V &= q^2 W_+ O_{2-}\\
        K_5^V &= 1/O_{2+}\\
        K_6^V &= 1/O_{2-}
        \end{aligned}$
        \end{tabular}
        \end{equation}
        where
        \begin{eqnarray}
        O_{1\pm} &=& \left [(W^2_{\pm} - k^2)(W^2_{\pm} - m_{\pi}^2)\right]^{\frac{1}{2}}/2W\nonumber\\
        O_{2\pm} &=& \left [(W^2_{\pm} - k^2)/(W^2_{\pm} - m_{\pi}^2)\right]^{\frac{1}{2}}.
        \end{eqnarray}
        \begin{table*}
        \centering
        \caption{Vector invariant amplitudes.}
        \label{inv_amp}
        \renewcommand{\arraystretch}{1.3}
        \begin{ruledtabular}
         \begin{tabular}{lclcl}
         Amplitude ~~~~~~ $e + p \rightarrow  e p \pi^{0}$ & ~~~~~~$ e + p \rightarrow e n \pi^{+}$    \\ [3pt]
         \hline
         $\begin{aligned}
          &\scalebox{0.01}{~}\\
         V_1~~~~~~~~&\frac{Mg_A}{f_{\pi}} \left(\frac{1}{s-M^2} + \frac{1}{u-M^2}\right)F_1(k^2) + \frac{g_A}{Mf_{\pi}} \mu_V F_2(k^2)\\
         V_2~~~~~~~~&\frac{Mg_A}{f_{\pi}} ~\frac{1}{qk}~\left(\frac{1}{s-M^2} + \frac{1}{u-M^2}\right)F_1(k^2)\\
         V_3~~~~~-&\frac{g_A}{2f_{\pi}} \left( \frac{1}{s-M^2} - \frac{1}{u-M^2}\right)F_2(k^2) \\
         V_4~~~~~-&\frac{g_A}{2f_{\pi}} \left( \frac{1}{s-M^2} + \frac{1}{u-M^2}\right)F_2(k^2)\\[6pt]
         V_5~~~~~0.\\[10pt]
         \end{aligned}$
         &
         $\begin{aligned}
          &\scalebox{0.01}{~}\\
         &\sqrt{2}\frac{Mg_A}{f_{\pi}} \left(\frac{1}{s-M^2} - \frac{1}{u-M^2}\right)F_1(k^2)\\
         &\sqrt{2}\frac{Mg_A}{f_{\pi}} ~\frac{1}{qk}~\left(\frac{1}{s-M^2} - \frac{1}{u-M^2}\right)F_1(k^2)\\
        -&\frac{g_A}{\sqrt{2}f_{\pi}} \left( \frac{1}{s-M^2} + \frac{1}{u-M^2}\right)F_2(k^2) \\
        -&\frac{g_A}{\sqrt{2}f_{\pi}} \left( \frac{1}{s-M^2} - \frac{1}{u-M^2}\right)F_2(k^2) \\
         -&\sqrt{2}\frac{g_A}{f_{\pi}}~\frac{1}{qk}~ \frac{1}{t- m_{\pi}^2} F_1(k^2)\\[3pt]
         \end{aligned}$
        \\
        \end{tabular}
        \end{ruledtabular}
        \end{table*}

        \nocite{*}
        
        \bibliography{bib_ma28}
        
        \end{document}